\documentclass[fleqn,12pt]{wlscirep}
\usepackage[utf8]{inputenc}
\usepackage[T1]{fontenc}
\usepackage[english]{babel}
\usepackage{amsmath}
\usepackage{amsfonts}
\usepackage{mathtools}

\usepackage{graphicx}
\usepackage{hyperref}
\usepackage[label font={bf, normalsize}]{subcaption}     
\usepackage{float}
\usepackage{color,soul}

\newcommand{\tsize}{\footnotesize} 

\usepackage{colortbl}
\usepackage{dcolumn}
\usepackage{bm}

\title{Scientific mobility, prestige and skill alignment in academic institutions}

\author[a†]{M\'arcia R. Ferreira}
\author[bc†]{R. Costas}
\author[a]{ V.D.P.~Servedio}
\author[ade*]{Stefan Thurner}
\affil[a]{Complexity Science Hub Vienna, Josefst\"adter Strasse 39, A-1080 Vienna, Austria}
\affil[b]{Centre for Science and Technology Studies, Leiden University, 2333 BN Leiden Leiden, Netherlands}
\affil[c]{Department of Science and Innovation-National Research Foundation of South Africa Centre of Excellence in Scientometrics and Science Technology Innovation Policy (SciSTP), Stellenbosch University, Stellenbosch, South Africa}
\affil[d]{Section for Science of Complex Systems, CeMSIIS, Medical University of Vienna, Spitalgasse 23, A-1090, Austria}
\affil[e]{Santa Fe Institute, 1399 Hyde Park Road, Santa Fe, NM 85701, United States}
\affil[*]{ \textbf{ferreira@csh.ac.at}} 
\affil[$†$]{†\textbf{Equal contribution}} 

\begin{abstract}
Scientific institutions play a crucial role in driving intellectual, social, and technological progress. Their capacity to innovate depends mainly on their ability to attract, retain, and nurture scientific talent and ultimately make it available to other organizations, industries, or the economy. As researchers change institutions during their careers, their skills are also transferred. The extent and mechanisms by which academic institutions manage their internal portfolio of scientific skills by attracting and sending researchers are far from being understood. We examine 25 million publication histories of 9.2 million scientists extracted from a large-scale bibliographic database covering thousands of research institutions worldwide to understand how the skills of mobile scientists align with those present in-house. We find a clear association between top-ranked institutions and greater skill alignment, i.e., the degree to which skills of incoming academics match those of their colleagues at the institution. We uncover similar high-alignment for scientists leaving top-ranked institutions. This type of academic alignment is more pronounced in engineering and life, health, earth, and physical sciences than in mathematics, computer science, social sciences, and the humanities. We show that over the past two decades, institutions generally have become more closely aligned in their overall skill profiles. We interpret these results in terms of levels of proactive management of the composition of the scientific workforce, diversity, and internal collaboration strategies at the institutional level.
\end{abstract}

\begin{document}

\flushbottom
\maketitle
%
%
\thispagestyle{empty}


\section*{Introduction}
Scientific discovery requires the capacity to seek, nurture, and combine internal and external sources of knowledge. Universities, in particular, serve as vital ``containers'' for the advancement and integration of this knowledge. However, because of the ``tacit'' nature of knowledge \cite{gertler2003tacit}, knowledge synergies do not emerge automatically. 
They rely on transfer mechanisms such as collaboration, networks, and labor mobility~\cite{cohen1990absorptive, zucker1994intellectual, gertler2003tacit, winter1982evolutionary}. Academic mobility is a particularly important mechanism for knowledge to flow effectively across people, organizations, locations, and time \cite{cohen1990absorptive, stephan2001exceptional, ganguli2015immigration}. 

Scientists are  moving between different institutions with increasing frequency \cite{sugimoto2017scientists}. According to some estimates, in 1990, about 2\% of scientists worked outside their country of origin \cite{stephan2001exceptional}. By 2000, this proportion increased to 14\% \cite{stephan2001exceptional}, and by 2015, it was estimated that about one-third of scientists were working outside their country of origin \cite{national2015revisiting}. A similar trend has been observed in Europe, where it has been reported that 7\% of hired researchers were from abroad~\cite{schiermeier2011career}. However, the presence of mobile researchers varies considerably by region and institution~\cite{machavcek2022researchers,schiermeier2011career, sugimoto2017scientists}. At Cambridge University, for example, it has been reported that more than 40\% of the faculty were foreign-born~\cite{schiermeier2011career}.  

\begin{figure} [t] 
\centering
\includegraphics[width=0.9\linewidth]{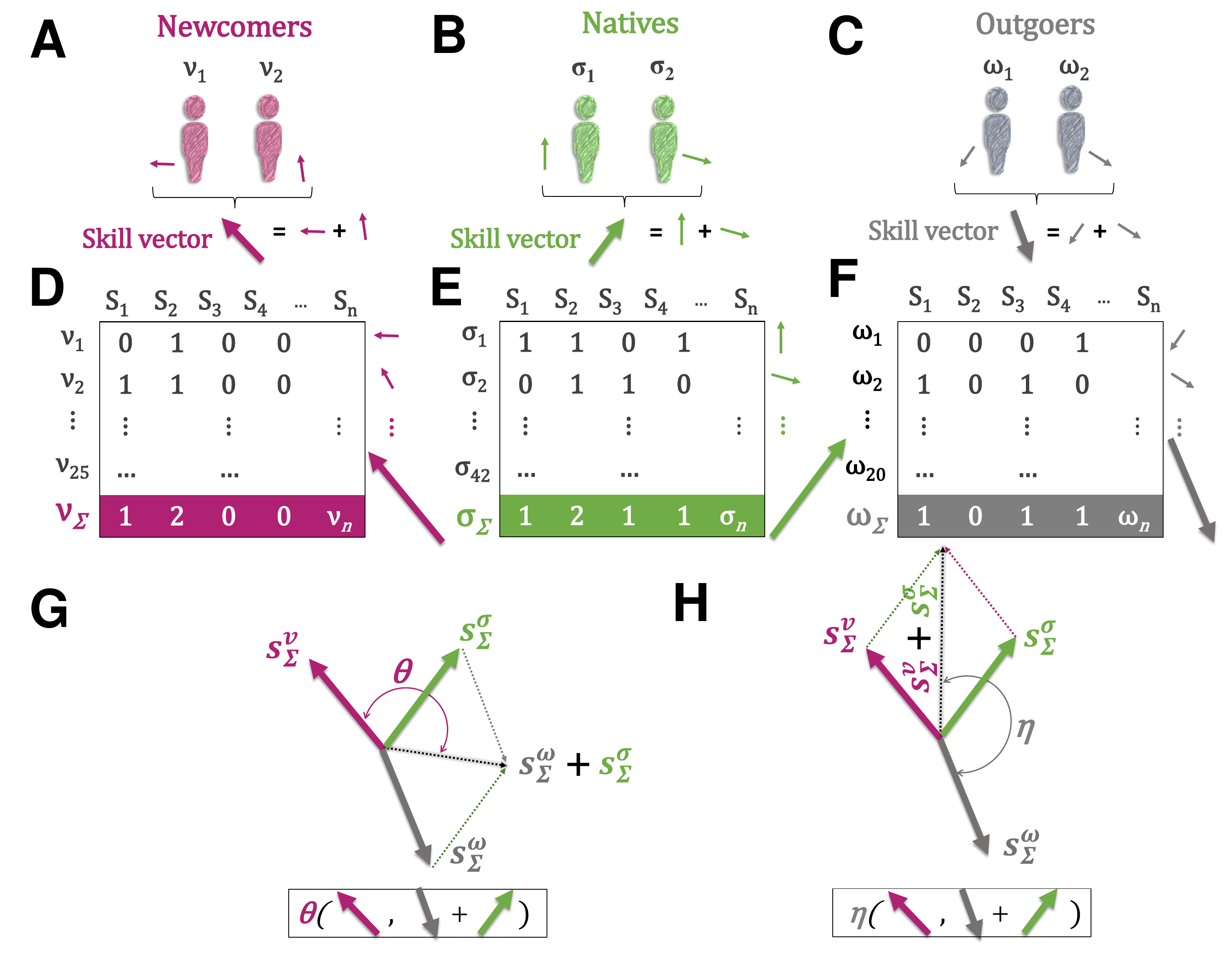}
\caption{
    A graphical representation of the skill and workforce structures of a scientific institution. We study three types of scientists (A, B, and C): \textit{institutional newcomers} $\nu$ (A), \textit{institutional natives} $\sigma$ (B), and \textit{institutional outgoers} $\omega$ (C). Every individual has a skill vector; every component in it, $S_k$, represents a particular skill $k$. We aggregate the individual skill vectors for each population (A), (B) and (C) of researchers and compute the cosine similarities between them, as shown in the grey boxes in panels G and H. We denote these similarities by $\theta$ (G) and $\eta$ (H). For illustration purposes, we took angles larger than $90\deg$. For definitions of the measures of skill alignment, see ~\hyperref[Materials and Methods]{Materials and Methods}.}
\label{Fig1:figure1abcdefgh_method.pdf}
\end{figure}

The attraction of mobile individuals to institutions has been studied for many decades \cite{grant1996toward,stephan2001exceptional, ganguli2015immigration}, and several analyses have shown that external talent is essential for innovation \cite{cohen1990absorptive, stephan2001exceptional, stephan2012economics, sugimoto2017scientists, ganguli2015immigration}. Attracting individuals trained in various research contexts is critical for frontier research \cite{ganguli2015immigration,stephan2001exceptional, sugimoto2017scientists, lepori2015competition, franzoni2014mover, milojevic2018changing, jaffe1996flows}, as it enables institutions to explore new areas of knowledge. However, it is also a challenge faced by most research institutions worldwide \cite{lepori2015competition}. The exchange of talent  is increasingly concentrated in a handful of universities~\cite{stephan2012economics}. In the United States, for example, the most prestigious institutions attracted and trained most of the available faculty before sending them to other mid- and upper-level research institutions \cite{clauset2015systematic,deville2014career}. Education systems also differ dramatically, with more prominent, well-funded universities offering more facilities, funding opportunities, and research diversity than smaller, specialized universities \cite{lepori2015competition,alvarez2021funding,costas2012approaching,lariviere2015team}. This unequal access to knowledge has significant implications for knowledge sharing across academic institutions and, more importantly, within the organization \cite{horn2007ranking,horta2009global}.

It has been argued that most academic institutions pursue the overarching goal of profile continuity and that the long-term sustainability of institutions can only be achieved through various forms of alignment \cite{heinze2008sponsor, march1991exploration}. Knowledge institutions typically invest and strengthen knowledge in their established research areas over time \cite{thurner2020role}, as leveraging on existing competencies can create alignments that improve performance, learning, and knowledge transfer~\cite{arthur1984competing, cohen1990absorptive}. When scientists within the institution have a common knowledge base, they can better learn from each other \cite{cohen1990absorptive}, leading to improved productivity and reduced barriers to collaboration. Yet, top institutions with large endowments are those that can experiment more in new emerging scientific fields.

An important driver of attracting talent that matches the internal profiles of institutions is the current reward system of science \cite{merton1957priorities}. This system often and increasingly discourages the pursuit of novel research areas because the returns from new ideas and topics are seen as uncertain, distant, and often risky \cite{heinze2008sponsor}. In contrast, the benefits of refining and expanding existing expertise and technologies are positive, immediate, and predictable \cite{march1991exploration}. Other studies have suggested that too much similarity in knowledge can also limit innovation \cite{fleming2001recombinant}, which at the institutional level means that as academic organizations exceed optimal levels of alignment, they potentially `lock' into dominant thematic profiles.

The mobility of academic talent, collaboration, and the alignment of skill profiles between institutions and incoming and outgoing researchers can play a critical role in shaping research dynamics within and across institutions. However, on a quantitative basis, there is limited understanding of the processes behind the alignment of knowledge and skills within institutions and the alignment of the skills of institutions and new hires. To gain a better understanding of these alignment strategies, we study academic institutions from the perspective of the composition of their workforce and the internal skill profiles they generate as the composition of their workforce changes over time. 

We use the {\em Dimensions} database (see ~\hyperref[Materials and Methods]{Materials and Methods}) to compare the internal skill profiles of millions of mobile individuals with the skill structures of the institutions they move to or leave. We quantify the academic skills of individuals by using their  publications mapped into a high-resolution classification scheme of scientific topics across all disciplines \cite{traag2019louvain}. This classification is the basis for defining the skill vector, $S^j$ for every individual, $j$. Every component of that binary vector represents a skill of the researcher, if the k-th component is $S^j_k=1$, researcher $j$ has competency $k$, if $S^j_k=0$, $j$ has no skill in $k$. If an author publishes in many different research areas, they has many `skills', if they publish on only one specific topic, the author has only a single non-zero component in the skill vector;  see ~\hyperref[Materials and Methods]{Materials and Methods}. The skills of an institution are defined as the superposition (sum of all vectors) of all the members of the existing workforce. These aggregated vectors are indicated as $S_\Sigma$, with a subscript $\Sigma$. The {\em Dimensions} database allows not only to quantify skills but also to observe the flows of researchers around the globe. 

However, measuring scientific skill profiles by bibliographic means is not an easy task. This partly depends on the level of resolution we use to determine researchers' skills. Data limitations have also been an obstacle to the study of scientists' knowledge pathways \cite{robinson2019many,sugimoto2017scientists, machavcek2022researchers}, leading to a prevalence of findings from self-reported information, small-scale studies, or studies limited to researchers from specific fields or countries \cite{jia2017quantifying,aleta2019explore, ganguli2015immigration, stephan2001exceptional, petersen2018multiscale, morgan2018prestige}. The situation is particularly problematic at the institutional level, as it relies on clear institutional identification and robust author-name disambiguation algorithms~\cite{donner2020comparing, machavcek2022researchers}. This situation has changed recently as more databases improve author and affiliation metadata \cite{robinson2019many,sugimoto2017scientists, machavcek2022researchers}. In what follows, we focus on harmonized research-intensive institutions data~\cite{hook2018dimensions,bode2018guide} for which extensive metadata on author-affiliation transitions exists~\cite{machavcek2022researchers}. 

In Fig.~\ref{Fig1:figure1abcdefgh_method.pdf}, we present a schematic view of how we approach the problem of skill assignment. We define three types of researchers: \textit{Newcomers} (A), \textit{Natives} (B), and \textit{Outgoers} (C). The natives represent the non-mobile workforce at a given institution, $i$. In the figure, we represent them as two scientists, $\sigma_1$ and $\sigma_2$, (green), both of which have different skills that are given by a skill vector, $S$, that has $n=4,163$ components that mark the different individual categories in the science classification scheme.  Native scientist $\sigma_1$ has three skills $S_1$, $S_2$, and $S_4$, hence $S^{\sigma_1}_1= S^{\sigma_1}_4= 1$, whereas $\sigma_2$ has only two, $S_2$ and $S_3$, $S^{\sigma_2}_2= S^{\sigma_2}_3= 1$, all other components being zero. Their combined skills are given by the sum of their skill vectors indicated by the small arrows. The skills present at the institution are seen in panel E. In this example, there are $r=42$ native researchers present; their skills are collected in the table. The sum of all their skills is called $S^{\sigma}_{\Sigma}$ and represents the current skill vector of the institution, $i$.  We now assume that in the next time period, a set of researchers will join the institution (newcomers, $\nu$) (A), and some will leave (the outgoers, $\omega$) (C). The collective skill vectors of these groups are called $S^{\nu}_{\Sigma}$ and $S^{\omega}_{\Sigma}$, respectively. In this example, we have 25 newcomers and 20 outgoers, with their skills captured in the tables in D and F. Data shows that the native population and newcomers make up the largest fraction in most institutions, while the outgoing population makes up the smallest fraction.

With this notion, we can now quantify the \emph{newcomer skill alignment} between the natives (plus outgoers) and the incoming workforce as the cosine of the angle, $\theta$, between the incoming skill vector, $S^{\nu}_{\Sigma}$ and the sum of natives and outgoers, $S^{\sigma}_{\Sigma}+S^{\omega}_{\Sigma}$; see panel (G). With this measure, we can analyze whether internally trained authors (natives and outgoers) and external authors (newcomers) generate aligned or divergent skill profiles at the institutional level. Similarly, we define the \emph{outgoer skill alignment} by calculating the cosine of the angle, $\eta$,  between the skill vector of outgoers, $S^{\omega}_{\Sigma}$ and the combined skill vectors of natives and newcomers, $S^{\sigma}_{\Sigma} + S^{\nu}_{\Sigma}$; see panel (H). 

\section*{Results}
\subsubsection*{Skill Alignment in Research Institutions}

\begin{figure} [!hbt] 
  \centering
  \includegraphics[width=0.8\linewidth]{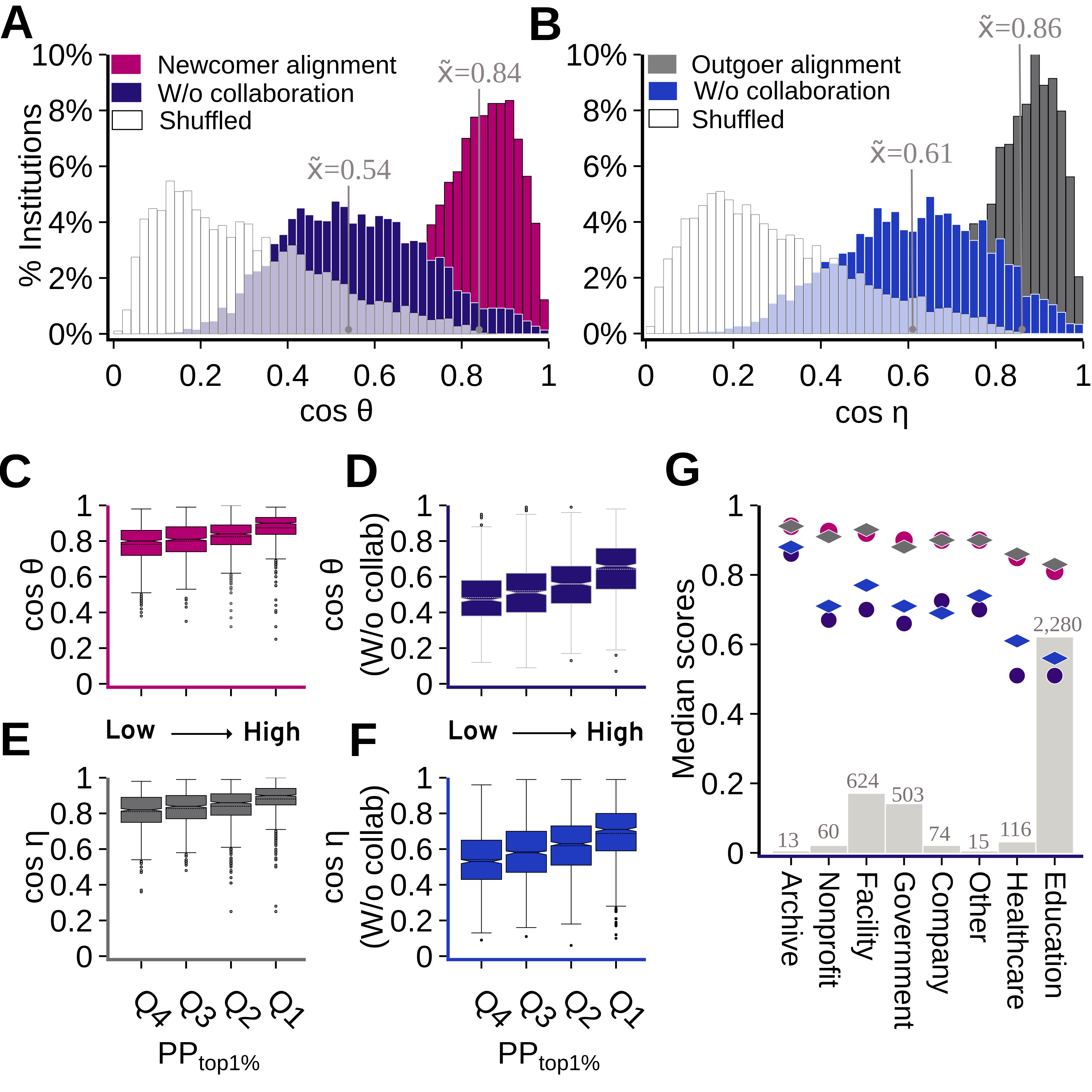}
  \caption{Alignments of the skill profile of the native faculty and newcomers to- and outgoers from all academic institutions. Panels A and B show the distributions for newcomers, $\cos{\theta}$, and outgoers, $\cos{\eta}$, respectively. 
  The purple (panel A) and blue (panel B) distributions show the alignment of skills between those newcomers and outgoers that were not collaborating with their peers at the institution before they joined or during their stay, respectively. The transparent lines represent a reference distribution of skill alignments obtained by shuffling the target (source) affiliations of newcomers (outgoers) ($n=10,000$ random assignments). Clearly, skill similarity is absent in the shuffled data. Panels C, D, E, and F show the influence of the institution's reputation. The alignment is shown for the quartiles within the top 1\% most impactful institutions, {PP}$_{{\rm top}1\%}$. The more impact, the more alignment, regardless of existing collaborations (compare C E and D E). Panel G captures the influence of institution type.
  It gives the median alignment scores by institution type. The shaded bars represent the percentage and number of institutions by organization type in our sample.~\label{fig2:figure2abcdefg_new.png}}
\end{figure}

To what extent does the skills profile of externally trained incoming scientists match that of institutional natives? Do their skills align, or are they different? Figures~\ref{fig2:figure2abcdefg_new.png}A and B show the cosine similarity between the skills profile of the institutions and its newcomer and outgoing workforce. We find a substantial similarity with a median of $0.84$ and $0.86$ for the newcomers and outgoers, respectively. The fact that the skill alignment is slightly lower for the newcomers than for the outgoers suggests that the outgoers have become more similar in their skills while they stayed at the institutions. Panels A (purple) and B (light blue) also show the similarity between the existing workforce skills and the skills profile generated by those newcomers and outgoers who did not interact with the rest of the institution's workforce during their stay. For these cases, we find much less similarity (median $0.54$ and $0.61$). This indicates that internal collaboration is a potential driver of intra-institutional skill alignment. The regression analysis shown in ~\hyperref[SI4]{SI text 2} confirms that collaboration within institutions is an important predictor of intra-institutional skill alignment.

To illustrate that the observed alignments are a significant and genuine effect that does not simply emerge as a statistical consequence of the definition of the cosine-similarity measure, we devise a simple ``null model''. We preserve the skill profiles of the in- and outgoers but remove the correlations with the profiles of the institutions. We do this by randomly assigning newcomer and outgoer skill profiles to institutions. The distributions are shown as transparent lines in Figures~\ref{fig2:figure2abcdefg_new.png}A and B. The skill alignments practically vanish as a result. 

There is a clear relation between institutional prestige, as captured by the {PP}$_\mathrm{top1\%}$ indicator (for definition, see ~\hyperref[Materials and Methods]{Materials and Methods}), and skill alignment. Figures~\ref{fig2:figure2abcdefg_new.png}C and E show that institutions that have substantially more than one percent of their publications in the top 1\% most cited papers worldwide tend to have similar skill profiles across the different types of workforce. Newcomers who move to an institution with top-cited publications tend to have more similar skill profiles than newcomers who move to a less prestigious institution, see C. The situation is similar for departing scientists, see E. In other words, talent flowing to and from organizations with high institutional prestige is associated with greater skill alignment. A greater skill dispersion is also observed at universities with lower prestige. If we compare C with D and E with F, we see that the prestige effect is independent of whether there are collaboration ties (i.e., co-authorship) between newcomers and outgoers with local researchers.

\begin{figure} [!hbt] 
  \centering
  \includegraphics[width=0.7\linewidth]{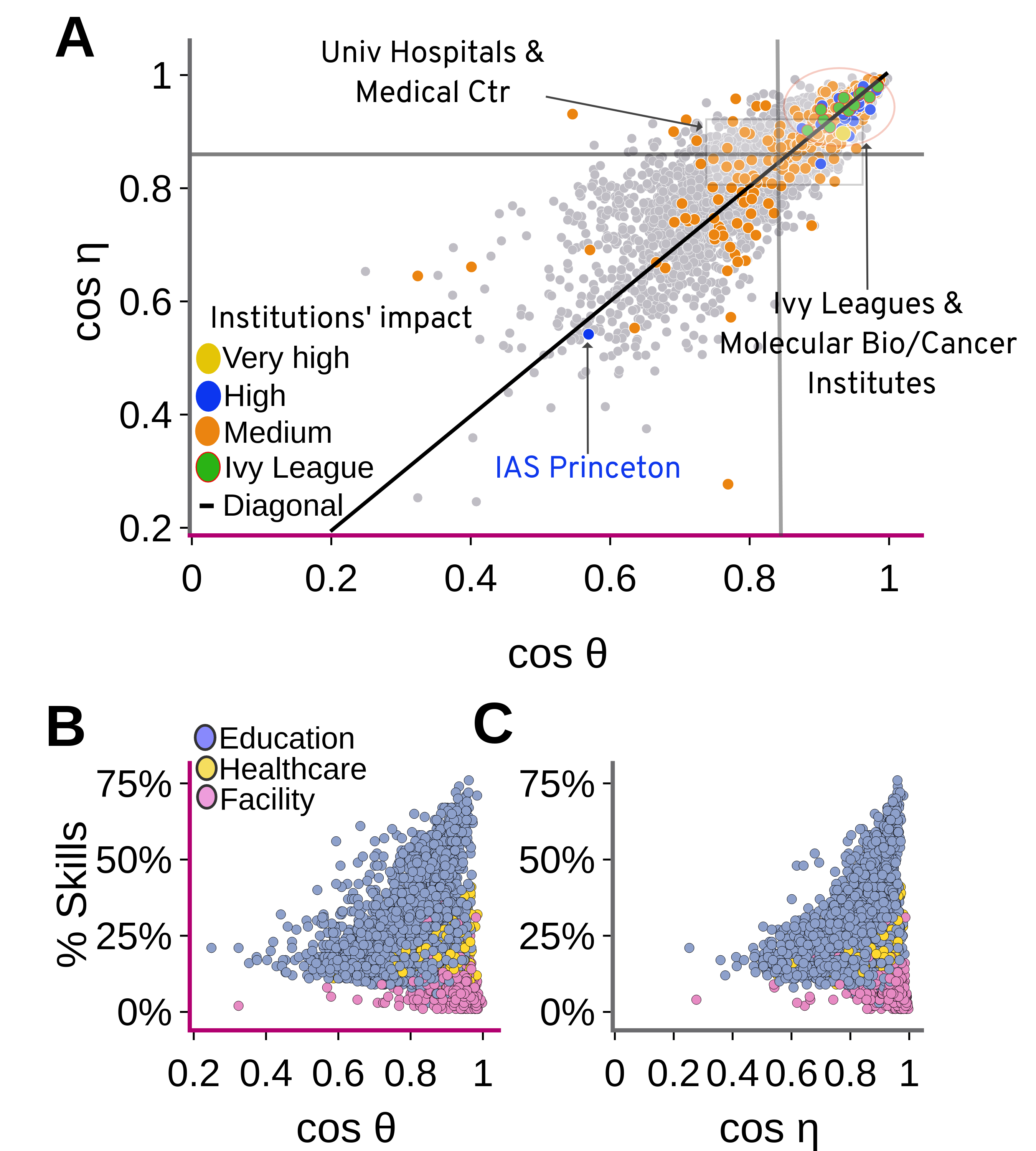}
  \caption{Scatter plot of the alignments of newcomers, $\cos{\theta}$, and outgoers, $\cos{\eta}$ (A). The line represents the diagonal (same in- and outgoer alignment). 
  The grey quadrant lines are positioned at the median values of $\cos{\theta}$ ($\tilde{x}=0.84$) and $\cos{\eta}$ ($\tilde{x}=0.86$). Every circle represents an institution. Color indicates the scientific impact of the institutions (${ PP }_\mathrm{top1\%}$ indicator). Grey institutions are below the global average (${ PP }_\mathrm{top1\%}\leq{0.01}$) of institutions with the same skills profile and years of production as explained in~\hyperref[Materials and Methods]{Materials and Methods}). Orange, blue, and yellow circles represent institutions with medium ($0.01\leq{ PP }_\mathrm{top1\%}\leq{0.05}$), high ($0.06\leq{ PP }_\mathrm{top1\%}\leq{0.09}$), and very high impact (${ PP }_\mathrm{top1\%}\geq{0.10}$), respectively. Top institutions tend to have generally high alignments and a slightly higher out-alignment. The scatter plots in panels B and C show the relation between the number of skills present at an institution (as a \% of all skills in the sample) and the skill alignments $\cos{\theta}$ (B) and $\cos{\eta}$ (C) for Education, Healthcare, and Facility research institutions, respectively. Healthcare and Facilities tend to have high in- and out-alignments; see also~\hyperref[SI]{SI}~Figure~\ref{SI:figure3abc_wo_collaboration.png}.}
  \label{fig3:figure3abc.png}
\end{figure}

Figure~\ref{fig2:figure2abcdefg_new.png}G shows the median alignment between newcomers and outgoers (colors correspond to those in panels A and B) and institutional natives by organization type. More generalist educational institutions (e.g., universities) tend to have lower median levels of similarity than more thematically focused institutions such as Facilities, Archives, Companies, Non-profits, or Governmental institutions. Interestingly, Healthcare research institutions also show comparatively low median scores of alignment, which may indicate that while they are considered specialized, their skill sets are broad enough to encompass a greater diversity of skill profiles between in and outgoing researchers and natives.

Figure~\ref{fig3:figure3abc.png}A shows the alignment of newcomers (x-axis) versus the alignment of outgoers (y-axis). The color indicates the degree of the citation's impact of institutions. The solid line marks the regression result. We segment the plot into four quadrants (gray lines at the median alignment values) associated with strategic patterns of talent attraction and training. Institutions (41\%) in the first quadrant (top right) attract and send the same skills at a rate above the median. Most notably, the U.S.\ Ivy Leagues, top European universities, and prominent molecular biology and cancer research institutes are in the first quadrant. There we also find several university hospitals and medical centers. This indicates that the institutions' strategy in the first quadrant is thematic continuity \cite{heinze2008sponsor, march1991exploration} and homogeneity in their recruitment and training practices. 

In the third quadrant (bottom left), we observe the opposite trend for about 45\% of institutions. Here, the profiles of newcomer hiring and outgoing researchers within an institution diverge and fall below the overall median scores. An example of this pattern is the Institute for Advanced Study (IAS) at Princeton (blue circle). This institute has a remarkably low alignment between outgoers and the rest of the institution, as well as between newcomers and the rest of the institution. Historically, the IAS has been a place where scientists retreat for sabbaticals and exchange ideas, encouraging unexpected discoveries and interdisciplinary thinking. This suggests that it is not always necessary to have a high-skill alignment of newcomers and outgoers to conduct high-impact research. Our method captures their (non) alignment strategy. However, IAS is an exception since this strategy seems prevalent at most institutions whose citation performance is below or close to the global average (grey).

The second quadrant (upper left) shows institutions (8\%) that attract researchers working in potentially complementary research areas. These institutions send researchers who produce skill-aligned profiles above the median of the other institutions to the institution's facilities and attract those who produce differentiated work below the median to the rest of the institution. In the fourth quadrant (bottom right), we find the opposite situation; about 6\% of institutions bring more of the same skills and send more researchers who exhibit different skills.

\begin{figure}[!hbt] 
\centering
\includegraphics[width=0.65\linewidth]{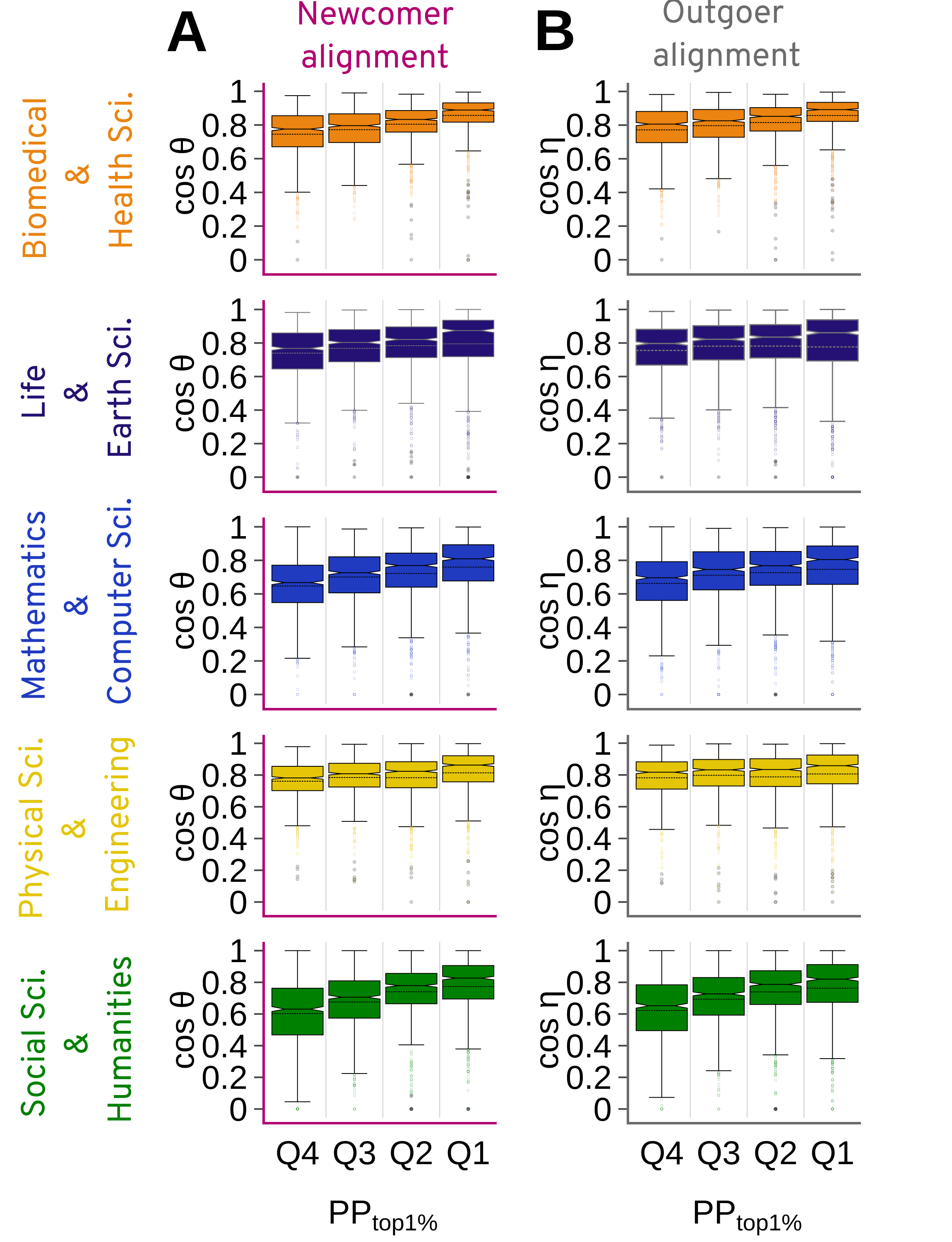}
\caption{Skill alignment of newcomers (A) and outgoers (B) 
in five main scientific fields as  a function of  institutional prestige as captured by impact quartiles of PP$_\mathrm{top1\%}$ (same figure setup as in Fig \ref{fig2:figure2abcdefg_new.png} C and E). Again, top-cited institutions have higher newcomer and outgoer alignment than average organizations (two-sample $t$ test, $P$ value <0.001). Disciplines are arranged alphabetically from top to bottom.
Social sciences, humanities, mathematics, and computer science show lower levels of alignment than biomedical and health sciences, life and earth sciences, physical sciences and engineering.
For the case controlled for collaborations, see~\hyperref[SI]{SI}~Figure~\ref{SI:figure8ab.png}.}
~\label{fig4:figure4ab.png}
\end{figure}

In Figures~\ref{fig3:figure3abc.png}B and C show the number of skills present at an institution (in \% of all possible 4,163 skills in our classification, see  see ~\hyperref[Materials and Methods]{Materials and Methods}), versus the skill alignment at institutions, B newcomers, C outgoers. We find considerable heterogeneity. Colors highlight three types of organizations: Education, Healthcare, and Facility, which account for 93\% of the institutions in our sample. The education category includes general and specialized universities, while the healthcare category includes university hospitals and medical research centers. Facilities, typically  established by the government or academic stakeholders, often specialize in one particular field, such as agriculture, high-energy physics, specific technologies, and others. We see that research facilities tend to be more specialized (small percentage of skills) and have higher skill alignments of both their incoming (B) and outgoing (C) workforce. Educational institutions tend to have a larger number of skills and show a large spread in both number of skills and alignment. Healthcare institutions fall between the two regarding skill diversity and show relatively high alignment values.

In~\hyperref[SI4]{SI text 2}, we conduct a multivariate regression analysis that examines the relationship between alignment and scientific impact measures and various controls while considering the different sizes of institutions. Our findings indicate that the level of internal collaboration within an institution and citation impact are important factors in determining skill alignment within academic institutions. 

\subsubsection*{Skill Alignment in Different Science Fields}
Various degrees of skill alignment are found in the five major areas of science. Figure~\ref{fig4:figure4ab.png} shows a breakdown of the distribution of alignment scores for newcomers (A) and outgoers (B). We find that the profiles for academics in the social sciences, humanities, mathematics, and computer science show lower levels of alignment. This is especially true for lower impact institutions, as captured by the quartiles of the proportion of papers in the top one percent (PP$_\mathrm{top1\%}$ indicator). Finally, in the fields of biomedical and health sciences, life and earth sciences, and physical sciences and engineering, there is a comparatively slightly higher degree of similarity between the skills of newcomers with their institution as well as between outgoers and the rest of their institution.

\subsubsection*{Skill Alignment Between Academic Institutions Over Time}
Finally, we present the situation of the alignment of skills between all the institutions in the sample. We analyze the  inter-institutional skill alignment, i.e., the similarity of skill profiles of the entire workforce of institutions between all institutions in the sample. The inter-institutional skill alignment is defined as the skill profile similarity denoted by $\cos{\phi}$ (see ~\hyperref[Materials and Methods]{Materials and Methods} for the definition). We compute it for all pairs of institutions.

\begin{figure} [!hbt] 
\centering
\includegraphics[width=0.4\linewidth]{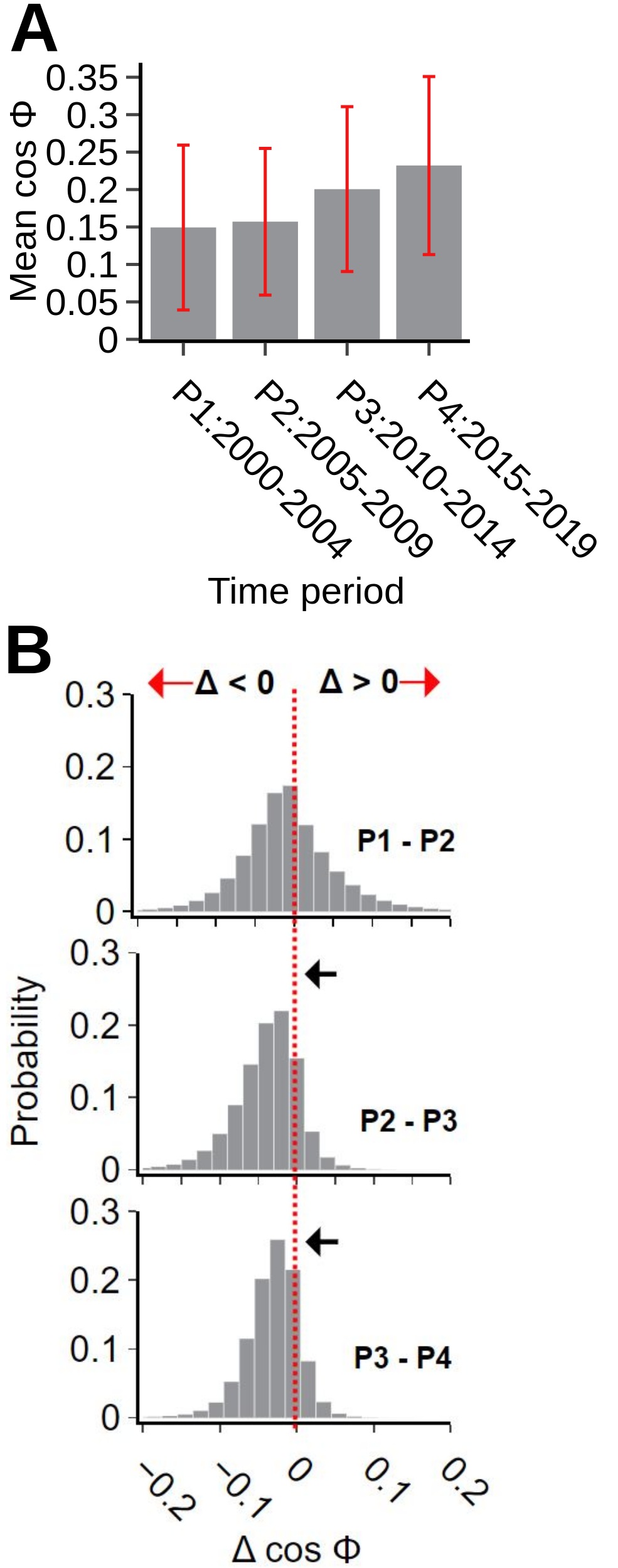}
\caption{Inter-institutional skill alignment over time. Panel A shows the average of pairwise cosine similarity for all pairs of institutions for four non-overlapping time periods: P1:2000-2004, P2:2005-2009, P3:2010-2014, and P4:2015-2019. The red error bars represent standard errors. Panel B shows the differences between the overall skill alignment of pairs of institutions, $\Delta\cos{\phi}$, over time. $\Delta<0$ means that institutions become increasingly similar in the composition of their overall skill profiles. The sub-panels capture the changes between the time periods. The tendency of becoming more similar is visible (over time, the peak is moving to the left).}
\label{fig5:figure5ab.PNG}
\end{figure}

Figure~\ref{fig5:figure5ab.PNG}A shows average inter-institutional skill alignment during four time periods from 2000 to 2019. The red error bars mark standard errors of the mean. We see that the average skill alignment is generally low, however, it doubled in the past twenty years, i.e., across the globe, skill profiles of institutions have become more similar. 

Figure~\ref{fig5:figure5ab.PNG}B shows the distribution of the differences in skill alignment between the pairs of institutions over time. A value of $\Delta \cos{\phi}$  below zero means that institutions have become more similar; when $\Delta \cos{\phi} >0$, institutions become more dissimilar. The dashed red line indicates the zero line of the x-axis. The three sub-panels capture the changes between the time periods, P1-P2, P2-P3, and P3-P4. It is visible that the peak of the distribution is moving toward the leftover time, indicating an acceleration toward becoming more similar in more recent years.

\section*{Discussion}
By quantifying the alignment of skills present at $3,965$ institutions, which includes the publication records of $9,299,250$ disambiguated authors affiliated with $108$ countries, with the $4,163$ skill types of the incoming and outgoing workforce, we can show strong quantitative signatures of academic skill alignment -- the degree to which mobile scholars (newcomers or outgoers) publish on topics that are in line with those of their colleagues already at the new institution. In particular, newcomers tend to publish on topics that align with those of their colleagues already at the new institution. Alignment, as measured by skills profile similarity, is more pronounced at the most prestigious (i.e., top-cited) institutions than at average institutions. Even within the top 1\% of highly cited institutions, there is a correlation between skill alignment and institutional citation performance. Research institutions with moderate levels of citation impact tend to have significantly less aligned skill profiles between natives and in- and outgoers. The greater alignment of skill profiles at top institutions is not surprising, as it indicates a strategic, specific, and targeted hiring policy that may not be present at more moderate institutions.

Highly aligned skill profiles potentially realize synergies between newcomers and existing faculty \cite{arthur1984competing, cohen1990absorptive} and can reinforce already strong research portfolios. However, this also may lead to selection pressures for those hired and the hiring institutions themselves and eventually lead to the under-representation of relevant research expertise, and important topics \cite{evans2014attention}. 

Two likely mechanisms could explain the origin of the observed similarities. One is that the newly hired scholars adapt their publication behavior (and scholarly interests) to the existing academic interests of the new institution. In this work, we see evidence that this may be the case, the outgoing researchers are more similar to the natives than the incoming researchers. This is reflected in a shift toward higher alignment and a narrower alignment distribution among outgoers. The other mechanism is the preference of institutions to hire scholars with similar skills to their current knowledge base or the preference of scholars to move to universities that are established in their fields. Also, a preferential dynamics of researchers going to places where lots of expertise exists, as, e.g. described in \cite{vitomarcia22}, might explain part of the observed effects. 

We also assessed the role of collaboration within the institution. We found a significant difference in the skill alignments of newcomers who have collaborative relationships (co-authorship on publications) with colleagues (natives or outgoers) at hiring institutions. Our results suggest that collaboration is the most natural approach for newcomers and natives to align, combine, and complement their skills at the institution. Both newcomers and outgoers who did not engage in collaborations are virtually unmatched by the local workforce, i.e., the cosine $\sim 0.5$. Note that skill differences between natives and newcomers are small when the degree of cooperation between them is high. The role of collaboration in skill alignment may also explain the disciplinary differences observed in our study, where Social Sciences, Humanities, Mathematics, and Computer Science show systematically lower levels of skill alignment. These disciplines traditionally have lower collaboration rates than the Natural Sciences or Engineering \cite{gazni2012mapping}.

A shortcoming of the present work is that individual preferences and motivations for mobility cannot be assessed. It would be interesting to supplement these results in future work with appropriately designed surveys and controlled experiments to uncover the relative importance of the individual-level mechanisms that lead to the observed profile alignments at the institutional level.

The presented results on the extent of skill alignments in scientific hiring can be considered steps toward a better understanding of talent flows in science. Future research would be important to determine how the alignment of institutions' skills profile interacts with different dimensions of workforce diversity, such as gender and seniority. This could provide policymakers with analytical tools to uncover the latent capabilities of different kinds of newcomers and assess their ability to influence (or not) the skill profile of their institutions.

Quantitative measures, such as those presented here, can inform and evaluate university (and unit) policy regarding their mobility, recruitment, and talent acquisition strategies, particularly about their existing competency profiles and those desired in the future. University leadership, funding agencies, and science policymakers in general-- may benefit from a quantitative assessment of the degree of alignment within their respective areas or organizations and can use it to develop interventions aimed at reaching desired alignment levels (e.g., by promoting internal collaboration networks).

\section*{Materials and Methods\label{Materials and Methods}}

Investigating the alignment between the skills profile of mobile scientists (newcomers or outgoers) and that of resident faculty at the institutional level requires data that describes the skills of individuals and captures the temporal information of the affiliations of every scientist. Such information is typically unavailable in surveys on a country's labor force. Even if available, the categories of highly skilled workers are often too ambiguous to identify specific groups of scientists \cite{stephan2001exceptional}. Therefore, we use data from the {\em Dimensions}\footnote{{\em Dimensions} is produced by {\em Digital Science} and was launched in January 2018. For more references, see \href{https://www.dimensions.ai/}{Dimensions.ai website}} database, which we accessed through the Centre for Science and Technology Studies (CWTS) at Leiden University. {\em Dimensions} covers local journals more comprehensively than other large-scale bibliographic databases such as {\em Web of Science} or {\em Scopus}. Its broader scope allows our analysis to be more inclusive of organizations with a more local focus, and thus, we also reduce mainstream effects \cite{machavcek2022researchers, hook2018dimensions}.

We examine the publication patterns of disambiguated authors from 108 countries between 2000 and 2020. Our analysis relies on three major improvements to the data: algorithmic disambiguation of author-names, improved consistency of organizations' metadata \cite{hook2018dimensions}, and a highly detailed field classification system \cite{traag2019louvain}, the latter also provided by the CWTS. We focus on disambiguated publications by authors with harmonized affiliation links. High-precision author disambiguation and institutional harmonization allow us to track the publication history of individual scientists across research institutions \cite{machavcek2022researchers}. This provides us with 9,299,250 million disambiguated author names.

Individual authors were disambiguated using the author-name disambiguation algorithm developed by {\em Dimensions} \cite{hook2018dimensions}, which uses the public ORCID\footnote{For more information, see \href{https://info.orcid.org/documentation/}{ORCID documentation}} as the basis for validating each author and their publication history. The disambiguation of organization names is based on the GRID\footnote{For more information, see \href{https://www.grid.ac/}{GRID website}} system \cite{bode2018guide}. For these authors, we retrieve their affiliations and publication history. We only consider publication-intensive institutions with at least 2,000 indexed publications, resulting in 3,965 institutions. Associated with these authors are 25,310,742 distinct documents indexed in the {\em Dimensions} database. The disambiguation of author names and the harmonization of research organization procedures allow us to produce a consistent overview of changes in researchers' affiliations with different institutions \cite{machavcek2022researchers} and topics in large-scale bibliometric analysis.

\subsection*{Measures of skill alignment for scientific institutions}
\label{Definition of skills section}
The skill sets of an institution's workforce are defined using the publication-level classification system of science developed by Waltman and van Eck \cite{waltman2012new}. The classification is done using the Leiden algorithm 
\cite{traag2019louvain} that clusters publications based on direct citation relations. With this method, we obtain a  detailed classification system for scientific literature that covers all scientific fields. It provides several features. First, it identifies the relatedness between pairs of 38.4 million publications indexed in {\em Dimensions} that are directly linked to 513 million citation relations. This step processes publications such as articles, reviews, book chapters, and proceedings from 2006 to 2020. In the second step, the publications are clustered into research areas using a clustering procedure and the areas are organized in a hierarchical structure \cite{traag2019louvain}. Finally, the methodology results in a hierarchical clustering system: i) a top level with 22 broad disciplines, ii) the second level with 824 areas, and iii) the third level with 4,163 micro-clusters. For more details on this approach, we refer to \cite{waltman2012new, traag2019louvain}.

In this paper, we consider the third classification level of 4,163 micro-clusters to define the skills profile vectors of institutions, as shown in Figure~\ref{Fig1:figure1abcdefgh_method.pdf}. In the figure, quantities with a subscript $\Sigma$ refer to the aggregate quantities of institutions. Formally, for an institution, $i$ (for which we have omitted the index $i$ in the figure for simplicity),
  \[ S^{\nu}_{\Sigma,i} = \sum_{r=1}^{N_i} S^{\nu_r}_{r,i} ~~,~~~ S^{\sigma}_{\Sigma,i} = \sum_{r=1}^{S_i} S^{\sigma_r}_{r,i} ~~,~~ S^{\omega}_{\Sigma,i} = \sum_{r=1}^{O_i} S^{\omega_r}_{r,i}
  \]\\
where $S^{\nu}_{r,i},\ S^{\sigma}_{r,i},\ S^{\omega}_{r,i}$ represent the $n$-component skill cluster vector of the newcomer, native, and outgoer scientists, $r$, respectively. The institution where these scientists are hosted is labeled by the index, $i$.
The values $N_i,\ S_i,\ O_i$ represent the total number of newcomers, natives, and outgoers, respectively, in institution, $i$. 
Components in the skill vectors are always binary, $1$ if the skill is present, $0$ if it is not present in an individual or at the institutional level. 

The angles $\theta_i$ and $\eta_i$ are defined according to the cosine similarity expressions with the Euclidean dot product:
\begin{equation} \label{Eq.~1}
 \cos \theta_i = 
    \frac{S^{\nu}_{\Sigma,i}\cdot(S^{\omega}_{\Sigma,i}+S^{\sigma}_{\Sigma,i})}
        {|S^{\nu}_{\Sigma,i}|\, |S^{\omega}_{\Sigma,i}+S^{\sigma}_{\Sigma,i}|}. 
\end{equation}
\begin{equation} \label{Eq.~2}
   \cos\eta_i = 
        \frac{S^{\omega}_{\Sigma,i}\cdot(S^{\nu}_{\Sigma,i}+S^{\sigma}_{\Sigma,i})}
            {|S^{\omega}_{\Sigma,i}|\, |S^{\nu}_{\Sigma,i}+S^{\sigma}_{\Sigma,i}|}.
\end{equation}
These definitions quantify the skill profile alignment of newcomers, $\cos\theta_i$, and the skills profile alignment of outgoers, $\cos\eta_i$, relative to the remaining researchers at academic institutions. A value of $\cos\theta_i = 1$ (or $\cos\eta_i = 1$) indicates that two vectors of skill clusters are identical for a given institution, reflecting the fact that the institution attracts new scientists and retains native scientists or promotes outgoing scientists and retains native scientists with the same micro-clusters or `skills', but also that there are scientists producing publication outputs that express these skills with the same weight. Conversely, a value of $0$ means that these different types of scientists do not have the same skills profile. 

We introduce a reference or ``null'' model for these two measures that retains the size of the institutions in terms of their total number of competencies and the number of authors. This is done to remove correlations between newcomers (outgoers) and the profile of the natives. This way, we recalculate $\cos\theta$ and $\cos\eta$ by randomly assigning newcomers and outgoers to institutions. This procedure allows us to disentangle actual ''local'' matching between scientists from statistical effects due to collaborative activities and the institutional size; see Figure~\ref{fig2:figure2abcdefg_new.png}A and B.

Using the same workforce components shown in Figure~\ref{Fig1:figure1abcdefgh_method.pdf}, we additionally calculate a measure of inter-institutional skill alignment between institution $i$ and $j$, $\cos\phi_{ij}(t)$, to track whether institutions' profiles are aligning or diverging over time. 
Here, $\phi_{ij}(t)$ is the angle between total (native, incoming, and outgoing) skill vectors of institutions $i$ and $j$ at time period, $t$ (i.e., 2000-2004, 2005-2009, 2010-2014, and 2015-2019).
This indicator estimates and accounts for an institution's entire workforce (i.e., no distinction is made between newcomers, natives, and outgoers) and reflects overall profile alignment across all pairs of institutions over four non-overlapping time periods, $t$ . 

We define inter-institutional alignment, $\cos\phi_{ij}(t)$, as
\begin{equation} \label{Eq.~3}
 \cos \phi_{ij}(t) = \frac{T_i\cdot T_j}{|T_i|\,|T_j|} \, .
\end{equation}\\
where $T_i=S^{\nu}_{\Sigma,i}+S^{\omega}_{\Sigma,i}+S^{\sigma}_{\Sigma,i}$ is the combined total skill vector of all scientists at institution, $i$. Finally, we capture the change in pairwise institutional alignment, which we refer to as inter-institutional alignment, as
\begin{equation} \label{Eq.~4}
 \Delta \cos \phi_{ij}(t) = \cos\phi_{ij}(t+1) - \cos\phi_{ij}(t) \, .
\end{equation}

Clustering relatively homogeneous publication sets into high-resolution clusters allows us to compare the aggregate capabilities of scientists within and across institutions. As we explain in the following section, this allows us to compute the citation impact of organizations in a similar research context \cite{waltman2012new, ruiz2015field}.

\subsection*{Citation impact indicators and normalization}
\label{impact indicators}
In recent years, in scientometrics, several changes took place that continue to influence the formal analyses of scientific dynamics. A growing awareness of the need to account for differences across and within disciplines when assessing the impact of research has increased research toward innovative indicators. In particular, field-normalized indicators based on bibliometric analyzes have become increasingly important for evaluating citation impact. For example, the average number of citations per publication varies significantly across scientific fields, institutions, and countries. The average number of citations per publication also varies by the age of the publication \cite{waltman2011towards}. Older publications are cited more frequently than more recent ones \cite{reisz2022loss}. Because of this uneven distribution of citations across different fields or years, citation counts or averages cannot be compared across research units \cite{waltman2011towards}. This is also important for the life and earth sciences, biomedical and health sciences, physical sciences and engineering, mathematics and computer science, and social sciences and humanities because these fields encompass different sub-disciplines and the sub-disciplines vary widely.

Taking these issues into account, we use the same high-resolution micro-clusters of topics used to define institutional competency vectors in \ref{Definition of skills section} - or the third level of classification and denoted by $S$ in \ref{Fig1:figure1abcdefgh_method.pdf} - to calculate the normalized citation indicators for each institution in our sample. These micro-clusters contain publications from multiple years (2000-2020), and each publication is assigned to a cluster based only on its citation relationships with other publications \cite{traag2019louvain}. We use a full-counting approach at the institutional level to calculate citation impact. That is, if two institutions contribute co-authors to a publication, the publication is counted as a full publication for both institutions. Using a full-counting approach, we give more weight to collaborative publications than non-collaborative ones~\cite{waltman2012new}. We use a publication window from 2006 to 2020 and a fixed citation window of four years to count citations to these papers through 2020. The authors' self-citations are not included in the calculation of impact indicators.

Before proceeding with the formal definition of citation-based indicators, we first consider a set of $n$ publications denoted by $1, \cdots, n$. Let $c_i$ denote the number of citations of a publication $i$, and $e_i$ denote the expected number of citations of publication $i$ given micro-cluster $S$ and year $t$ in which publication $i$ was published. In other words, $e_i$ is the average number of citations of all publications published in the same micro-cluster and year as publication $i$. We define two indicators of citation impact for each research institution: the {\em total normalized citation score}, TNCS, and the {\em proportion of publications in the top $n^{th}\%$}, PP$_\mathrm{top\,nth\%}$. The TNCS indicator captures an institution's total normalized citation rate of the produced publication volume. It is similar to what \cite{lundberg2007lifting} calls the total field normalized citation score indicator and is defined as
\begin{equation} \label{Eq.~5}
  \mathrm{TNCS} = \sum\nolimits_{i=1}^{n}\frac{c_i}{e_i} \, .
\end{equation}\\
The PP$_\mathrm{top\,nth\%}$ uses percentile rank classes instead of mean-based indicators to normalize the citation impact of publications \cite{bornmann2013use, waltman2012new, waltman2013calculation}. It measures the proportion of articles among the top $n^{th}$\% most cited papers in the same skill or micro-cluster, of the same age and document type. We first assign a percentile based on its position in the citation distribution of articles in the same micro-cluster. We use the approach described in \cite{waltman2013calculation} to calculate the percentile rank of each publication. In our analysis, we compute three variations of this metric. Specifically, we use the $99^\mathrm{th}$, the $95^{th}$ and the $90^\mathrm{th}$ percentile ranks, which assign papers with a percentile equal to or greater than the $99^\mathrm{{th}}$, $95^\mathrm{{th}}$, and $90^\mathrm{{th}}$ percentile to the top 1\%, 5\%, and 10\% of frequently cited papers, respectively. The percentile rank measures, PP$_\mathrm{top\,1\%}$, PP$_\mathrm{top\,5\%}$, and the PP$_\mathrm{top\,10\%}$ of each publication was calculated using
\begin{equation} \label{Eq.~6}
 \mathrm{ PP }_\mathrm{top\,x\%} = \frac{\sum_{i=0}^{\infty} n_i c_i} {\sum_{i=0}^{\infty}\ n_i} \, ,
\end{equation}
where $n_i$ denotes the number of publications from a scientific institution with $i$ citations. The score of a publication with $i$ citations is indicated by $c_i$. According to this definition, the PP$_\mathrm{top\,nth\%}$ is simply the average score of the publications of the unit of analysis. For simplicity, the definition assumes that all publications of a given unit belong to the same cluster. For more details, see \cite{waltman2013calculation, waltman2012leiden, waltman2011towards}.

\renewcommand\bibname{References}


\section*{Additional information}
\textbf{Competing interests}\\
All authors declare that they have no competing interests.\\

\noindent\textbf{Funding}\\
We acknowledge support from the Austrian Research Promotion Agency FFG under grants 857136 and 882184, as well as the South African DST-NRF Centre of Excellence in Scientometrics and Science, Technology and Innovation Policy (SciSTIP).\\

\noindent We also thank Giordano de Marzo, Frank Neffke and Niklas Reisz for fruitful discussions and the Centre for Science and Technology Studies (CWTS) at Leiden University for access to the {\em Dimensions} database.

\clearpage
\onecolumn

\section*{Supplementary information}

\renewcommand{\thesubsection}{\Alph{subsection}.}
\begin{figure} 
\centering
  \includegraphics[width=0.65\linewidth]{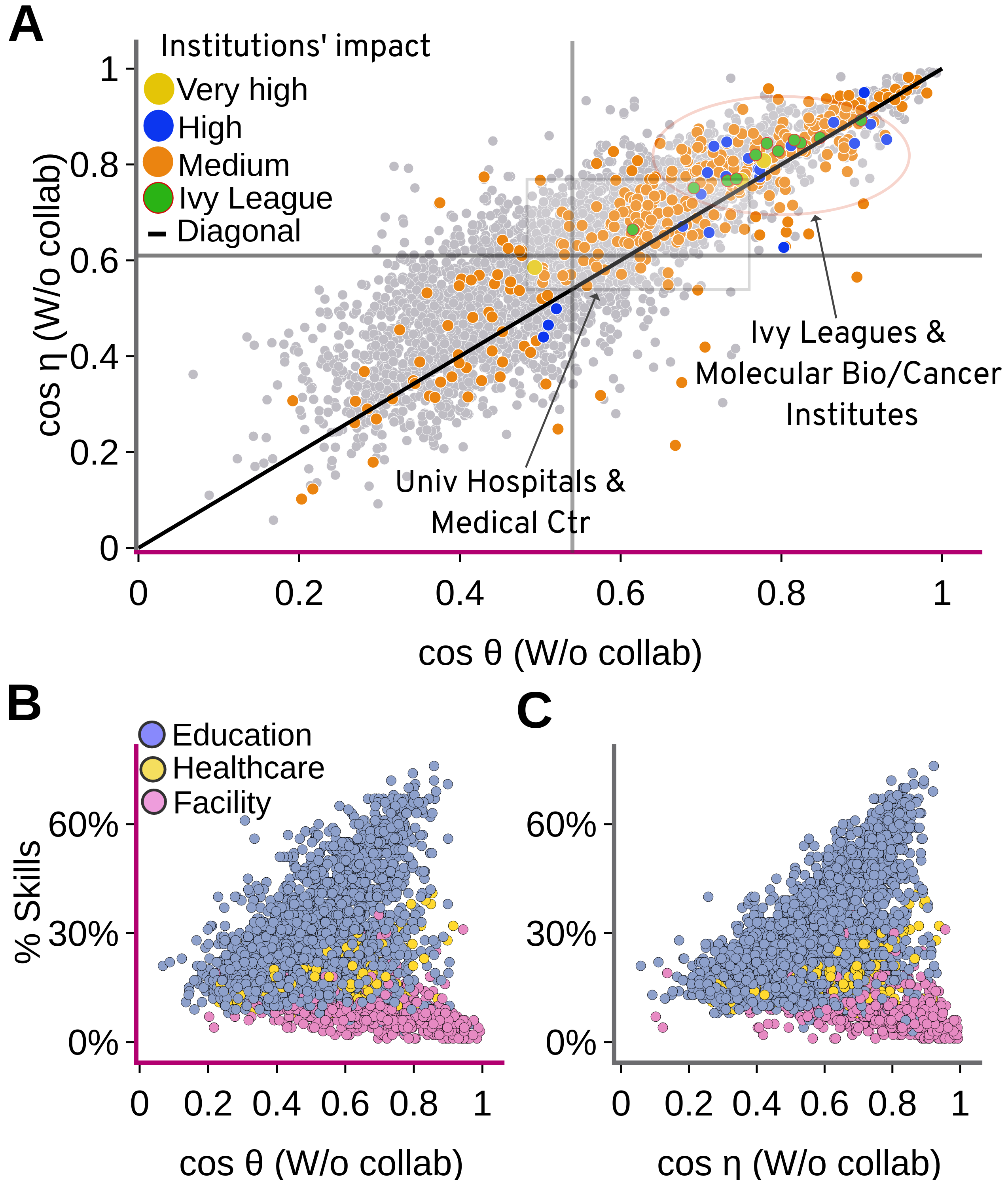}
\caption{Scatter plot of the correlation between newcomer $\cos{\theta}$ (W/o collab)- and outgoer $\cos{\eta}$ (W/o collab) alignment measures.Grey quadrant lines of median $\cos{\theta}$ (W/o collab) ($\tilde{x}=0.54$) and $\cos{\eta}$ (W/o collab) ($\tilde{x}=0.61$) are shown. Each circle in A represents an institution. The color of the circles represents the scientific impact of the institutions as captured by the PP$_\mathrm{top\,1\%}$ indicator. Grey-colored institutions have a PP$_\mathrm{top\,1\%}$ that is close to or below the global average (PP$_\mathrm{top\,1\%}\leq{0.01}$) of institutions with the same skill profile and years of production. The orange, blue, and yellow circles represent institutions with medium ($0.01\leq \mathrm{PP}_\mathrm{top\,1\%}\leq{0.05}$), high ($0.06\leq\mathrm{PP}_\mathrm{top\,1\%}\leq{0.09}$), and very high impact (PP$_\mathrm{top\,1\%}\geq{0.10}$). 35\% of institutions are in quadrant 1, 29\% in quadrant 2, 36\% in quadrant 3, and 1\% in quadrant 4. The scatter plots in panels B and C show the relationship between an institution's percentage of its total skills and $\cos{\theta}$ (W/o collab) and $\cos{\eta}$ (W/o collab), respectively, for education, health, and research facilities when articles in collaboration between the different types of the workforce are excluded. The panels have the same setting as the Figure~\ref{fig3:figure3abc.png}, but this time for the measures $\cos\theta$ (W/o collab) and $\cos\eta$ (W/o collab), i.e. for the case of non-collaborative hires and leavers. We find much greater variability than in the main measures of skill alignment in the main text. Similar to Figure~\ref{fig3:figure3abc.png}, the skill profile of newcomers and outgoers is concentrated in the first and third quadrants, followed by the second and fourth quadrants. Interestingly, ``Ivy League'' universities and molecular biology and cancer research institutes remain concentrated in the first quadrant. In contrast, university hospitals and medical research centres are mainly found in the second and first quadrants. The most frequently cited institutions are mostly concentrated around high scores for non-collaborative alignment in quadrant one, with some exceptions spreading across the other three quadrants. Panels B and C show a very similar trend compared to figures~\ref{fig3:figure3abc.png}B and C.
}
\label{SI:figure3abc_wo_collaboration.png}
\end{figure}

\begin{figure} 
\centering
  \includegraphics[width=0.6\linewidth]{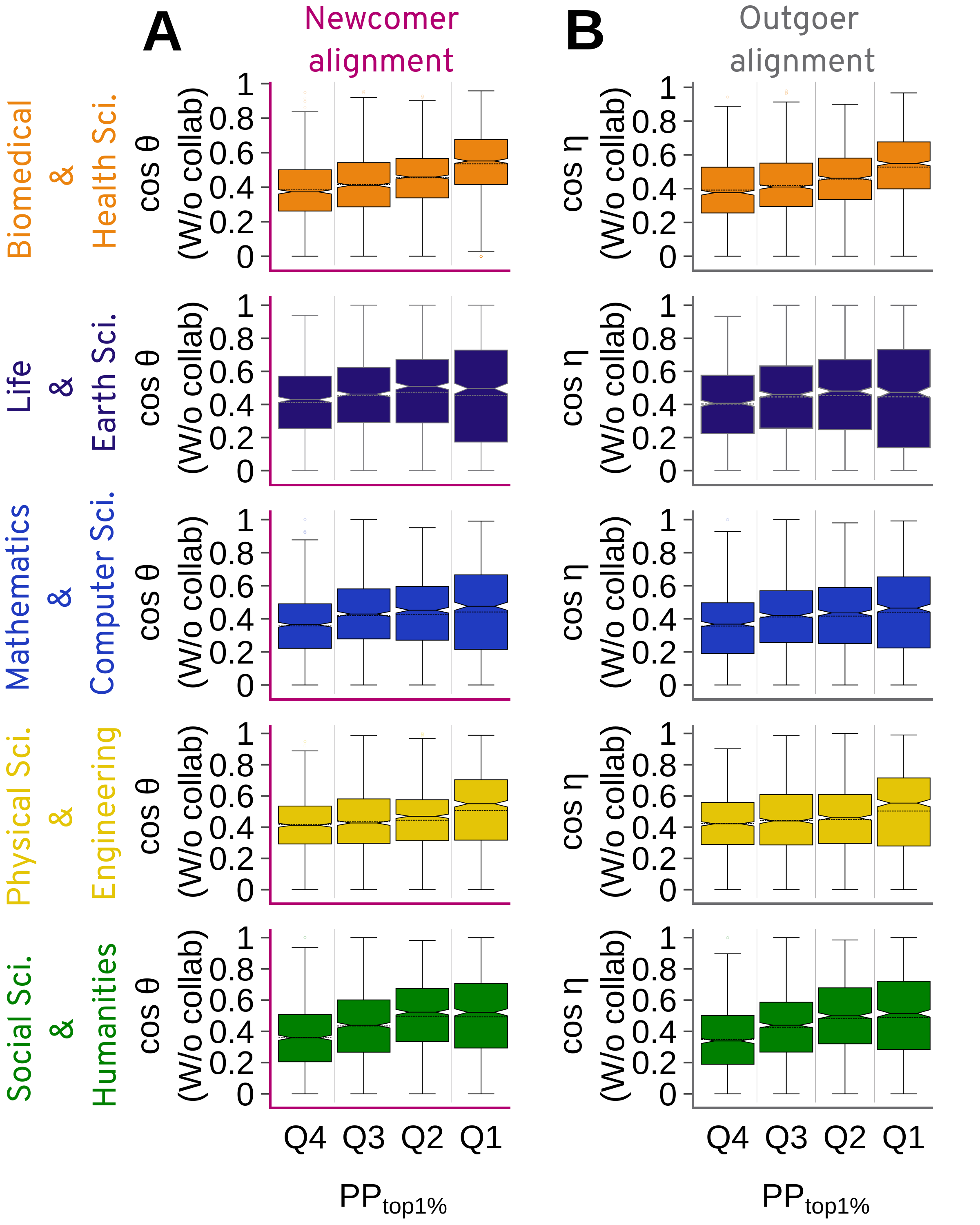}
\caption{
Boxplots of newcomer skill alignment, $\cos{\theta}$(W/o collab), and outgoer skill alignment, $\cos{\eta}$(W/o collab), by major fields of science and quartiles of PP$_{top1\%}$. We provide an overview of disciplinary trends in the alignment of mobile researchers' skills that occurs when scholars within an institution do not collaborate.
Top-cited institutions have higher alignment than average organizations (two-sample $t$ test, $P$ value <0.001) even when researchers do not collaborate internally. This finding can be generalized across large scientific fields and suggests that the alignment of capabilities that occurs from non-collaborative work positively affects the citation visibility of institutions. Here the quartiles from low impact (i.e., quartile 4) to high impact (i.e., quartile 1) and the disciplines are arranged alphabetically from top to bottom. The panels have the same setting as the Figure~\ref{fig4:figure4ab.png}, but this time for the $\cos\theta$ (W/o collab) and $\cos\eta$ (W/o collab) measures, i.e., for the case of non-collaborative hires and leavers. We find that the alignment of their skills is significantly lower across impact quartiles and disciplines. Nevertheless, alignment scores are higher for the most cited institutions (i.e., quartile 1) than for the least cited institutions (i.e., quartile 4). This tells us that internal collaboration is also an important factor in integrating knowledge and aligning skills in institutions and across the disciplines in which the institutions operate. Moreover, the most frequently cited institutions still attract and nurture similarly skilled researchers, even if they do not collaborate on scientific publications within the institution. 
}
\label{SI:figure8ab.png}
\end{figure}

\subsection*{SI Text 1: Classification of disambiguated authors into Newcomers, Natives, and Outgoers\label{SI2}} 

Researchers affiliated with each organization in the data are split into three categories based on where they began publishing in relation to the organization being analyzed: a) Natives -- are those who started publishing on the same organization and have not (yet) changed their affiliation, b) Newcomers -- are those who started publishing on a different organization, and c) Outgoers -- are those who left the first organization where they published at some point under the period of analysis (2000-2020). 

\subsection*{SI Text 2: Multivariate Linear Regression Analysis\label{SI4}} 
In this section, we delve into the details of our regression analysis. We first define several indicators to better understand the connection between skill alignment, intra-institutional collaboration, scientific impact, and productivity measures. To ensure the validity of our findings, we also conduct robustness checks to test whether our main findings still hold when alternative measures of scientific influence are used (see sections~\hyperref[SI4.1]{SI Text 2.1}, and ~\hyperref[SI4.2]{SI Text 2.2}).

\subsubsection*{SI Text 2.1: Regression of newcomer skill alignment\label{SI4.1}}
Our main results are based on a multivariate linear ordinary least squares regression model as below:

\begin{equation} \label{equation 7}
\begin{array}{lcl}
{\cos \theta}_{i} &=& \beta_0\\
&& + \beta_{1,2} {\mathrm{X}}_{i}\\
&& + \beta_{3} {\mathrm{r}}_{i}\\
&& + \beta_{4} {\mathrm{PP}_{\nu}}_{i}\\
&& + \beta_{5} {\mathrm{PPP}\,(\mathrm{collab}_{\nu})_{i}}\\
&& + \beta_{6} {\mathrm{PPS}_{i}: \mathrm{Diversified}}\\
&& + \epsilon_i \, .\\
\end{array}
\end{equation} 

\textbf{Dependent Variable.} The dependent variable  is a measure of how closely the skills of newcomers and the existing workforce at an institution, $i$, match. This measure is represented by $\cos{\theta}_i$ ,which is defined in~\hyperref[Definition of skills section]{Eq.~1}.  Additionally, we also use another measure, $\cos{\eta}_i$, for outgoers who have changed their institutional affiliations, which is defined in ~\hyperref[Definition of skills section]{Eq.~2}.\\ 

\textbf{Predictors of Interest.}The predictors of interest in this study are three percentile-based citation impact metrics (PP$_\mathrm{top,1\%i}$, PP$\mathrm{top,5\%i}$, and PP$\mathrm{top,10\%_i}$) and a continuous variable of normalized citation counts (TNCS$_i$). These measures will be used to explain the skill alignment of newcomers and existing workforce in an institution by using a regression model. For the term X$_i$ in Eq.~\ref{equation 7} we first use three percentile-based independent variables, PP$_\mathrm{top\,1\%_i}$, PP$_\mathrm{top\,5\%_i}$, and PP$_\mathrm{top\,10\%_i}$, which indicate the fractions of upper-tail papers being the top 1\%, 5\%, and 10\%, respectively, as measured by citation (see definition in ~\hyperref[Definition of skills section]{Eq.~6}). These measures are included separately in different regression models, with coefficient $\beta_1$. Finally, with coefficient $\beta_2$, we use an alternative measure of scientific impact in X$_i$, the continuous variable TNCS$_{i}$, defined in \hyperref[Definition of skills section]{Eq.~5}, which measures the total normalised citation count of a scientific institution.\\

\textbf{Control Variables.} In this study, several other explanatory variables are included as independent variables to control for other possible predictors of workforce skill alignment. These variables include size and internal collaboration indicators at the institutional level. Descriptive statistics on the variables used in the regression analysis are presented in \hyperref[Table S1]{Table S1}. The following control variables will be used in the regression analysis of newcomers' skill alignment presented in this section:

\newcommand\textequal{%
 \rule[.4ex]{4pt}{0.4pt}\llap{\rule[.7ex]{4pt}{0.4pt}}}

\begin{itemize} 
  \item r$_{i}$~\textequal~indicates the total number of researchers at a research institution 
  \item PP$_{\nu_i}$~\textequal~ indicates the proportion of newcomers of a research institution
  \item PPP (collab$_\nu)_{i}$~\textequal~indicates the proportion of publications in collaboration between newcomers and the rest of the institution
 \item PPS$_{i}$~\textequal~ is a dichotomous variable that indicates whether an institution has an overall diversified or specialized skills' portfolio. It indicates the absence or presence of diversification or specialization that may be expected to shift the outcome of skill alignment produced by the total workforce of a research institution. PPS$_{i}$ = 1 if the institution has produced an above median proportion of skills and PPS$_{i}$ = 0 otherwise
\end{itemize}

\noindent These control variables are used to account for other factors that may influence the skill alignment of newcomers and existing workforce in an institution. Furthermore, we included PPS$_{i}$, and r$_{i}$ in all regression models. All independent variables included in the regression models had a variance inflation score (VIF) well below four, below the tolerance threshold of 0.25.\\

\textbf{Stratification by Size of Institutions.} The total publication output, $P_{i}$, is used as an indicator of the size of an institution. Institutions' size has been shown to affect the ability of institutions to attract researchers \cite{machavcek2022researchers, clauset2015systematic, Zhang22}. It is therefore assumed that size also modifies the proportion of publications in the top $1\%$, $5\%$, and $10\%$ of an institution's publications on skill alignment. To control for this confounding effect, we stratified institutions into three subgroups based on their publication production, creating subgroups in which an institution's publication production varies less than all institutions in the sample combined. This stratification strategy will be applied in all regression models.\\

The main results of the stratified regression analysis are presented in ~\hyperref[Table S2]{Table S2},~\hyperref[Table S3]{Table S3}, and ~\hyperref[Table S4]{Table S4}. The coefficients of the tables are standardized. Standardizing the coefficients allows us to compare the strength of the effect of each individual independent variable on the dependent variable. This is important because size-dependent associations can produce substantially higher correlation coefficients, especially when they exhibit substantial variance \cite{traag2019systematic}. This can potentially alter the relative associations between alignment and normalized count-based citation measures. 

The results of models (1), (2), (3), and (4) in the three tables suggest that there is a significant relationship between an institution's skill alignment and the proportion of publications in the top 1\%, 5\%, 10\%, and Total Normalized Citation Score (TNCS) after controlling for several other explanatory variables. These variables include the proportion of incoming researchers, the proportion of internal collaboration between the newcomers and the rest of the institution's workforce, the total workforce, and the degree of diversification of skills. This relationship is consistent across all subgroups of institutional size. The findings also indicate that skill alignment is, on average, lower in diversified institutions than in specialized institutions, as indicated by the binary variable PPS$_{i}$: Diversification/Specialization, which was negative in all models and tables. In specialized institutions, which are more thematically focused, the expertise of mobile researchers better matches the expertise of their institutions, while in diversified institutions there are more opportunities for topical diversification. The proportion of incoming researchers, while significant, does not appear to be a particularly strong predictor of skill alignment.

Collaboration activity within the institution is one of the strongest predictors of skill alignment, with a standardized coefficient between 0.455 and 0.442 across different subgroups of institutional size. This suggests that institutions that encourage internal collaboration among researchers may be more successful in aligning the expertise of newcomers with that of the institution as a whole. The proportion of papers in the top 1\%, 5\% and 10\% of highly cited publications are also found to be significantly associated with skill alignment, with standardized coefficients of 0.168(0.026), 0.214(0.028) and 0.203(0.028) respectively in, for example,~\hyperref[Table S2]{Table S2}. For medium and high output institutions, total workforce capacity (r$_{i}$) is also found to be considerably associated with higher levels of alignment, with the coefficients ranging from 0.224(0.023) to 0.213(0.024) in~\hyperref[Table S3]{Table S3} and slightly higher for high output institutions, ranging from 0.307(0.048) to 0.288(0.022) in~\hyperref[Table S4]{Table S4}.

Overall, these results suggest that internal collaboration and citation impact likely explain the alignment at institutions, particularly in institutions with lower publication output, while medium and large publication output institutions also achieve greater alignment by attracting most of the available scientific labour~\cite{Zhang22}. The citation impact variables for medium and large institutions remain significantly associated with alignment, except for the total normalized citation score (TNCS$_{i}$) in model (4) in~\hyperref[Table S4]{Table S4} for high output institutions. This suggests that while publications at the high end of the citation percentile scale may play a role in aligning an institutions expertise, citation `quantity' does not have a direct effect on the alignment of skills among newcomers and the rest of the workforce in larger scientific institutions. 


\begin{table*}[!ht]
\caption*{Table S1. Descriptive statistics of the main indicators and control variables for all academic institutions in the sample (mean, median, standard deviation, skewness, and kurtosis).}
\label{Table S1}
\centering
\begin{tabular}{@{}
>{\columncolor[HTML]{FFFFFF}}c 
>{\columncolor[HTML]{FFFFFF}}c 
>{\columncolor[HTML]{FFFFFF}}c 
>{\columncolor[HTML]{FFFFFF}}c 
>{\columncolor[HTML]{FFFFFF}}l 
>{\columncolor[HTML]{FFFFFF}}l 
>{\columncolor[HTML]{FFFFFF}}l @{}}
\toprule
\multicolumn{7}{c}{\cellcolor[HTML]{FFFFFF}\textbf{Descriptive Statistics}}         \\ \midrule
N                          & N   & Mean & Median & \multicolumn{1}{c}{\cellcolor[HTML]{FFFFFF}Std. Deviation} & Skewness & Kurtosis \\ \midrule
PP$_{top1\%_i}$   & 3,965 & 0.02   & 0.01   & 0.01   & 2.25 & 11.58 \\ \midrule
PP$_{top5\%_i}$ & 3,965 & 0.07   & 0.06   & 0.03   & 1.35 & 3.77 \\ \midrule
PP$_{top10\%_i}$ & 3,965 & 0.13   & 0.12   & 0.05   & 0.97 & 1.69 \\ \midrule
TNCS$_{i}$  & 3,965 & 9,809.58 & 3,813.59 & 18,324.12 & 5.36 & 41.00 \\ \midrule
P$_{i}$      & 3,965 & 7,196.51 & 3,610.00 & 11,238.25 & 4.27 & 24.13 \\ \midrule
r$_{i}$    & 3,965 & 3,834.33 & 1,894.00 & 5,899.69 & 4.99 & 40.13 \\ \midrule
PP$_{\nu_i}$    & 3,965 & 0.41   & 0.39   & 0.17   & 0.37 & -0.55 \\ \midrule
PP$_{\omega_i}$      & 3,965 & 0.19   & 0.19   & 0.06   & 0.46 & 0.58 \\ \midrule
PPP (collab$_{\nu})_{i}$ & 3,965 & 0.30   & 0.30   & 0.10   & 0.40 & 1.68 \\ \midrule
PPP (collab$_{\omega})_{i}$ & 3,965 & 0.27  & 0.27   & 0.11   & 0.44 & 0.60 \\ \midrule
PPS$_{i}$  & 3,965 & 0.21  & 0.17 & 0.13 & 1.15 & 0.87 \\ \midrule
cos $\theta_i$        & 3,965 & 0.82 & 0.84  & 0.10                            & -0.99  & 1.41   \\ \midrule
cos $\eta_i$       & 3,965 & 0.84 & 0.86  & 0.10                            & -1.24  & 2.16   \\ \midrule
cos $\theta$ (W/o collab)$_{i}$ & 3,965 & 0.55 & 0.54  & 0.17                            & 0.19   & -0.45  \\ \midrule
cos $\eta$ (W/o collab)$_{i}$  & 3,965 & 0.60 & 0.61  & 0.17                            & -0.16  & -0.52  \\ \bottomrule
\end{tabular}
\end{table*}

\begin{table*}[!htbp] \centering 
  \caption*{Table S2. Relationship between low output institutions' citation impact and newcomer skill alignment. To allow comparison of the coefficients in the table, all variables were centred. Therefore, the beta coefficients in the table have standard deviations as their units. The results in models (1), (2), (3), and (4) show that the proportion of publications in the top 1\%, 5\%, 10\% and the continuous variable TNCS$_{i}$ have a significant association with the alignment between newcomers and the rest of the workforce. The degree of internal collaboration within institution, PPP(collab$_{\nu})_{i}$, is found to be most strongly associated with alignment, while the proportion of newcomers, PP$_{\nu_i}$, and the total workforce, r$_{i}$, are least strongly associated with skill alignment.}
  \label{Table S2} 
{\tsize
\begin{tabular}{@{\extracolsep{5pt}}lcccc} 
\\[-1.8ex]\hline 
\hline \\[-1.8ex] 
 & \multicolumn{3}{c}{\textit{Dependent variable: $\cos{\theta}_{i}$}} \\ 
\cline{2-5} 
\\[-1.8ex] & \multicolumn{4}{c}{Institutions with Low Publication Output, P$_{i}$ [387, 2425]} \\ 
\\[-1.8ex] & (1) & (2) & (3) & (4)\\ 
\hline \\[-1.8ex] 
 Constant & 0.086$^{***}$ (0.025) & 0.082$^{***}$ (0.025) & 0.080$^{***}$ (0.025) & 0.093$^{***}$ (0.025) \\ 
  r$_{i}$ & 0.094$^{***}$ (0.023) & 0.096$^{***}$ (0.023) & 0.098$^{***}$ (0.023) & 0.088$^{***}$ (0.023) \\ 
  PP$_{\nu_i}$ & 0.137$^{***}$ (0.025) & 0.106$^{***}$ (0.027) & 0.094$^{***}$ (0.028) & 0.143$^{***}$ (0.025) \\ 
  PPP (collab$_\nu)_{i}$ & 0.455$^{***}$ (0.022) & 0.446$^{***}$ (0.022) & 0.442$^{***}$ (0.022) & 0.443$^{***}$ (0.023) \\ 
  PP$_{{top\,1\%}_{i}}$ & 0.168$^{***}$ (0.026) &  &  &  \\ 
  PP$_{{top\,5\%}_{i}}$ &  & 0.203$^{***}$ (0.028) &  &  \\ 
  PP$_{{top\,10\%}_{i}}$ &  &  & 0.214$^{***}$ (0.028) &  \\ 
  TNCS$_{i}$ &  &  &  & 0.164$^{***}$ (0.025) \\ 
 \hline \\[-1.8ex] 
PPS$_{i}$: Diversification/Specialization dummy? & Yes & Yes & Yes & Yes \\ 
\hline \\[-1.8ex] 
Observations & 1,322 & 1,322 & 1,322 & 1,322 \\ 
Adjusted R$^{2}$ & 0.386 & 0.391 & 0.392 & 0.385 \\ 
Residual Std. Error (df = 1316) & 0.784 & 0.780 & 0.780 & 0.784 \\ 
F Statistic (df = 5; 1316) & 166.946$^{***}$ & 170.617$^{***}$ & 171.361$^{***}$ & 166.655$^{***}$ \\ 
\hline 
\hline \\[-1.8ex] 
Standard errors in parentheses. & \multicolumn{4}{l}{$^{*}$p$<$0.1; $^{**}$p$<$0.05; $^{***}$p$<$0.01} \\ 
\end{tabular} 
}
\end{table*} 

\begin{table*}[!htbp] \centering 
  \caption*{Table S3. Relationship between medium output institutions' citation impact and newcomer skill alignment. To allow comparison of the coefficients in the table, all variables were centred. Therefore, the beta coefficients in the table have standard deviations as their units. The results in models (1), (2), (3), and (4) show that the proportion of publications in the top 1\%, 5\%, 10\% and the continuous variable TNCS$_{i}$ have a significant association with the alignment between newcomers and the rest of the workforce. The degree of internal collaboration within institution, PPP(collab$_{\nu})_{i}$, is most strongly associated with alignment within institutions, followed by the total workforce, r$_{i}$, of the institution while the proportion of newcomers, PP$_{\nu_i}$ is least strongly associated with skill alignment.} 
  \label{Table S3} 
{\tsize 
\begin{tabular}{@{\extracolsep{5pt}}lcccc} \\[-1.8ex]\hline 
\hline \\[-1.8ex] 
 & \multicolumn{4}{c}{\textit{Dependent variable: $\cos{\theta}_{i}$}} \\ 
\cline{2-5} 
\\[-1.8ex] & \multicolumn{4}{c}{Institutions with Medium Publication Output, P$_{i}$ [2425, 4864]} \\ 
\\[-1.8ex] & (1) & (2) & (3) & (4)\\ 
\hline \\[-1.8ex] 
 Constant & 0.244$^{***}$ (0.031) & 0.248$^{***}$ (0.031) & 0.249$^{***}$ (0.031) & 0.257$^{***}$ (0.031) \\ 
  r$_{i}$ & 0.224$^{***}$ (0.023) & 0.222$^{***}$ (0.023) & 0.222$^{***}$ (0.023) & 0.213$^{***}$ (0.024) \\ 
  PP$_{\nu_i}$ & 0.073$^{***}$ (0.024) & 0.062$^{**}$ (0.024) & 0.059$^{**}$ (0.024) & 0.096$^{***}$ (0.024) \\ 
  PPP (collab$_\nu)_{i}$ & 0.431$^{***}$ (0.022) & 0.426$^{***}$ (0.022) & 0.423$^{***}$ (0.022) & 0.429$^{***}$ (0.022) \\ 
  PP$_{{top\,1\%}_{i}}$ & 0.182$^{***}$ (0.024) &  &  &  \\ 
  PP$_{{top\,5\%}_{i}}$ &  & 0.202$^{***}$ (0.024) &  &  \\ 
  PP$_{{top\,10\%}_{i}}$ &  &  & 0.208$^{***}$ (0.024) &  \\ 
  TNCS$_{i}$ &  &  &  & 0.140$^{***}$ (0.024) \\ 
 \hline \\[-1.8ex] 
PPS$_{i}$: Diversification/Specialization dummy? & Yes & Yes & Yes & Yes \\ 
\hline \\[-1.8ex] 
Observations & 1,321 & 1,321 & 1,321 & 1,321 \\ 
Adjusted R$^{2}$ & 0.406 & 0.412 & 0.414 & 0.396 \\ 
Residual Std. Error (df = 1315) & 0.770 & 0.767 & 0.766 & 0.777 \\ 
F Statistic (df = 5; 1315) & 181.735$^{***}$ & 186.102$^{***}$ & 187.440$^{***}$ & 174.110$^{***}$ \\ 
\hline 
\hline \\[-1.8ex] 
Standard errors in parentheses. & \multicolumn{4}{l}{$^{*}$p$<$0.1; $^{**}$p$<$0.05; $^{***}$p$<$0.01} \\ 
\end{tabular} 
}
\end{table*} 

\begin{table*}[!htbp] \centering 
  \caption*{Table S4. Relationship between high output institutions' citation impact and newcomer skill alignment. To allow comparison of the coefficients in the table, all variables were centred. Therefore, the beta coefficients in the table have standard deviations as their units. The results in models (1), (2), (3), and (4) show that the proportion of publications in the top 1\%, 5\%, 10\% have a significant association with the alignment between newcomers and the rest of the workforce. The degree of internal collaboration within institution, PPP(collab$_{\nu})_{i}$, is most strongly associated with alignment within institutions, followed by the total workforce, r$_{i}$, of the institution while the proportion of newcomers, PP$_{\nu_i}$ is least strongly associated with skill alignment.} 
  \label{Table S4} 
{\tsize  
\begin{tabular}{@{\extracolsep{5pt}}lcccc} 
\\[-1.8ex]\hline 
\hline \\[-1.8ex] 
 & \multicolumn{4}{c}{\textit{Dependent variable: $\cos{\theta}_{i}$}} \\ 
\cline{2-5} 
\\[-1.8ex] & \multicolumn{4}{c}{Institutions with High Publication Output, P$_{i}$ [4864, 121750]} \\ 
\\[-1.8ex] & (1) & (2) & (3) & (4)\\ 
\hline \\[-1.8ex] 
 Constant & 0.588$^{***}$ (0.086) & 0.584$^{***}$ (0.086) & 0.588$^{***}$ (0.086) & 0.693$^{***}$ (0.085) \\ 
  r$_{i}$ & 0.295$^{***}$ (0.022) & 0.288$^{***}$ (0.022) & 0.288$^{***}$ (0.022) & 0.307$^{***}$ (0.048) \\ 
  PP$_{\nu_i}$ & 0.070$^{***}$ (0.027) & 0.057$^{**}$ (0.027) & 0.058$^{**}$ (0.027) & 0.162$^{***}$ (0.022) \\ 
  PPP (collab$_\nu)_{i}$ & 0.429$^{***}$ (0.021) & 0.425$^{***}$ (0.021) & 0.422$^{***}$ (0.021) & 0.424$^{***}$ (0.022) \\ 
  PP$_{{top\,1\%}_{i}}$ & 0.159$^{***}$ (0.028) &  &  &  \\ 
  PP$_{{top\,5\%}_{i}}$ &  & 0.178$^{***}$ (0.029) &  &  \\ 
  PP$_{{top\,10\%}_{i}}$ &  &  & 0.175$^{***}$ (0.029) &  \\ 
  TNCS$_{i}$ &  &  &  & 0.021 (0.048) \\ 
 \hline \\[-1.8ex] 
PPS$_{i}$: Diversification/Specialization dummy?  & Yes & Yes & Yes & Yes \\ 
\hline \\[-1.8ex] 
Observations & 1,322 & 1,322 & 1,322 & 1,322 \\ 
Adjusted R$^{2}$ & 0.430 & 0.432 & 0.432 & 0.416 \\ 
Residual Std. Error (df = 1316) & 0.755 & 0.754 & 0.754 & 0.764 \\ 
F Statistic (df = 5; 1316) & 199.974$^{***}$ & 202.121$^{***}$ & 201.545$^{***}$ & 188.853$^{***}$ \\ 
\hline 
\hline \\[-1.8ex] 
Standard errors in parentheses. & \multicolumn{4}{l}{$^{*}$p$<$0.1; $^{**}$p$<$0.05; $^{***}$p$<$0.01} \\ 
\end{tabular} 
}
\end{table*}

\subsubsection*{SI Text 2.2: Regression of outgoer skill alignment\label{SI4.2}}
The results in this section are based on a multivariate linear ordinary least squares regression model as below:

\begin{equation} \label{equation 9}
\begin{array}{lcl}
{\cos \eta}_{i} &=& \beta_0\\
&& + \beta_{1,2} {\mathrm{X}}_{i}\\
&& + \beta_{3} {\mathrm{r}}_{i}\\
&& + \beta_{4} {\mathrm{PP}_{\omega}}_{i}\\
&& + \beta_{5} {\mathrm{PPP}\,(\mathrm{collab}_{\omega})_{i}}\\
&& + \beta_{6} {\mathrm{PPS}_{i}: \mathrm{Diversified}}\\
&& + \epsilon_i \, .\\
\end{array}
\end{equation} 

\textbf{Dependent Variable.} The dependent variable is the skill alignment of the outgoer and remaining workforce of an institution, $i$, $\cos{\eta}_i$ defined in~\hyperref[Definition of skills section]{Eq.~2}. \\

\textbf{Predictors of Interest.} Just like in the previous section, we use, for the term X$_i$, three percentile-based independent variables, ${ PP }_{top1\%}$, ${ PP }_{top5\%}$, and ${ PP }_{top10\%}$, measures to indicate the fractions of upper-tail papers being the top 1\%, 5\%, and 10\%, respectively, gauged by citation (See definition in the~\hyperref[Definition of skills section]{Eq.~6}). These measures are included separately in different regression models with coefficient $\beta_1$. The total normalized citation counts of a scientific institution TNCS$_i$, defined in ~\hyperref[Definition of skills section]{Eq.~5} is also tested in X$_i$ with coefficient $\beta_2$.\\

\textbf{Control Variables.} In order to account for other potential factors that may impact outgoer skill alignment, we have included a number of explanatory variables in our analysis. These variables include measures of institutional size and internal collaboration, and the descriptive statistics for these variables can be found in~\hyperref[Table S1]{Table S1}. These variables will be used as control variables in our regression analysis.

\begin{itemize} 
  \item r$_{i}$~\textequal~indicates the total number of researchers at a research institution 
  \item PP$\omega_{i}$~\textequal~ indicates the proportion of outgoers of a research institution
  \item PPP (collab$_{\omega})_{i}$~\textequal~indicates the proportion of publications in collaboration between outgoers and natives of the institution
  \item PPS$_{i}$~\textequal~ is a dichotomous variable that indicates whether an institution has an overall diversified or specialized skills' portfolio. It indicates the absence or presence of diversification or specialization that may be expected to shift the outcome of skill alignment produced by the total workforce of a research institution. PPS = 1 if the institution has produced an above median proportion of skills and PPS = 0 otherwise 
\end{itemize}

These control variables are used to account for other factors that may influence the skill alignment of outgoers and existing workforce in an institution. Furthermore, we included PPS$_{i}$, and r$_{i}$ in all regression models. All independent variables included in the regression models had a variance inflation score (VIF) well under four, below the tolerance threshold of 0.25.\\

\textbf{Stratification by Size of Institutions.} The total publication output, $P_{i}$, is used as an indicator of the size of an institution. Institutions' size has been shown to affect the ability of institutions to attract researchers \cite{machavcek2022researchers, clauset2015systematic}. It is therefore assumed that size also modifies the proportion of publications in the top $1\%$, $5\%$, and $10\%$ of an institution's publications on outgoer skill alignment. Stratification allows us to control for this confounding effect by creating three subgroups in which an institution's publication production varies less than all institutions in our sample combined.\\

\begin{table*}[!htbp] \centering 
  \caption*{Table S5. Relationship between low output institutions' citation impact and outgoer skill alignment. To allow comparison of the coefficients in the table, all variables were centred. Therefore, the beta coefficients in the table have standard deviations as their units. The results in models (1), (2), (3), and (4) show that the proportion of publications in the top 1\%, 5\%, 10\% and the continuous variable TNCS$_{i}$ have a significant association with the alignment between outgoers and the rest of the workforce. The degree of internal collaboration within institution, PPP(collab$_{\omega})_{i}$, is most strongly associated with alignment within institutions, followed by the total workforce, r$_{i}$ of the institution while the proportion of newcomers, PP$_{\omega_i}$ and the total workforce, r$_{i}$, are least strongly associated with skill alignment.} 
  \label{Table S5} 
  \label{} 
{\tsize
\begin{tabular}{@{\extracolsep{5pt}}lcccc} 
\\[-1.8ex]\hline 
\hline \\[-1.8ex] 
 & \multicolumn{4}{c}{\textit{Dependent variable: $\cos{\eta}_{i}$}} \\ 
\cline{2-5} \\[-1.8ex] & \multicolumn{4}{c}{Institutions with Low Publication Output, P$_{i}$ [387, 2425]} \\ 
\\[-1.8ex] & (1) & (2) & (3) & (4)\\ 
\hline \\[-1.8ex] 
 Constant & 0.084$^{***}$ (0.023) & 0.077$^{***}$ (0.022) & 0.074$^{***}$ (0.022) & 0.094$^{***}$ (0.023) \\ 
  r$_{i}$ & 0.118$^{***}$ (0.022) & 0.114$^{***}$ (0.021) & 0.115$^{***}$ (0.021) & 0.113$^{***}$ (0.022) \\ 
  PP$_{\omega_i}$ & 0.166$^{***}$ (0.025) & 0.163$^{***}$ (0.024) & 0.160$^{***}$ (0.024) & 0.161$^{***}$ (0.025) \\ 
  PPP (collab$_\omega)_{i}$ & 0.471$^{***}$ (0.025) & 0.476$^{***}$ (0.025) & 0.478$^{***}$ (0.025) & 0.459$^{***}$ (0.025) \\ 
  PP$_{{top\,1\%}_{i}}$ & 0.212$^{***}$ (0.020) &  &  &  \\ 
  PP$_{{top\,5\%}_{i}}$ &  & 0.236$^{***}$ (0.020) &  &  \\ 
  PP$_{{top\,10\%}_{i}}$ &  &  & 0.243$^{***}$ (0.020) &  \\ 
  TNCS$_{i}$ &  &  &  & 0.205$^{***}$ (0.020) \\ 
 \hline \\[-1.8ex] 
PPS$_{i}$: Diversification/Specialization dummy? & Yes & Yes & Yes & Yes \\ 
\hline \\[-1.8ex] 
Observations & 1,322 & 1,322 & 1,322 & 1,322 \\ 
Adjusted R$^{2}$ & 0.484 & 0.493 & 0.496 & 0.482 \\ 
Residual Std. Error (df = 1316) & 0.718 & 0.712 & 0.710 & 0.720 \\ 
F Statistic (df = 5; 1316) & 248.656$^{***}$ & 258.183$^{***}$ & 260.884$^{***}$ & 246.584$^{***}$ \\ 
\hline 
\hline \\[-1.8ex] 
Standard errors in parentheses. & \multicolumn{4}{l}{$^{*}$p$<$0.1; $^{**}$p$<$0.05; $^{***}$p$<$0.01} \\ 
\end{tabular} 
}
\end{table*} 

\begin{table*}[!htbp] \centering 
  \caption*{Table S6. Relationship between high output institutions' citation impact and outgoer skill alignment. To allow comparison of the coefficients in the table, all variables were centred. Therefore, the beta coefficients in the table have standard deviations as their units. The results in models (1), (2), (3), and (4) show that the proportion of publications in the top 1\%, 5\%, 10\% have a significant association with the alignment between outgoers and the rest of the workforce. The degree of internal collaboration within institution, PPP(collab$_{\omega})_{i}$, is most strongly associated with alignment within institutions, followed by the total workforce, r$_{i}$, of the institution while the proportion of outgoers, PP$_{\omega_i}$, is least strongly associated with skill alignment.} 
  \label{Table S6} 
{\tsize 
\begin{tabular}{@{\extracolsep{5pt}}lcccc} 
\\[-1.8ex]\hline 
\hline \\[-1.8ex] 
 & \multicolumn{4}{c}{\textit{Dependent variable: $\cos{\eta}_{i}$}} \\ 
\cline{2-5} 
\\[-1.8ex] & \multicolumn{4}{c}{Institutions with Medium Publication Output, P$_{i}$ [2425, 4864]} \\ 
\\[-1.8ex] & (1) & (2) & (3) & (4)\\ 
\hline \\[-1.8ex] 
 Constant & 0.178$^{***}$ (0.028) & 0.179$^{***}$ (0.028) & 0.180$^{***}$ (0.028) & 0.197$^{***}$ (0.028) \\ 
  r$_{i}$ & 0.211$^{***}$ (0.022) & 0.206$^{***}$ (0.022) & 0.206$^{***}$ (0.022) & 0.199$^{***}$ (0.022) \\ 
  PP$_{\omega_i}$ & 0.121$^{***}$ (0.023) & 0.116$^{***}$ (0.023) & 0.116$^{***}$ (0.023) & 0.128$^{***}$ (0.024) \\ 
  PPP (collab$_\omega)_{i}$  & 0.502$^{***}$ (0.024) & 0.505$^{***}$ (0.024) & 0.503$^{***}$ (0.024) & 0.486$^{***}$ (0.024) \\ 
  PP$_{{top\,1\%}_{i}}$ & 0.209$^{***}$ (0.020) &  &  &  \\ 
  PP$_{{top\,5\%}_{i}}$ &  & 0.227$^{***}$ (0.020) &  &  \\ 
  PP$_{{top\,10\%}_{i}}$ &  &  & 0.228$^{***}$ (0.020) &  \\ 
  TNCS$_{i}$ &  &  &  & 0.182$^{***}$ (0.020) \\ 
 \hline \\[-1.8ex] 
PPS$_{i}$: Diversification/Specialization dummy? & Yes & Yes & Yes & Yes \\ 
\hline \\[-1.8ex] 
Observations & 1,321 & 1,321 & 1,321 & 1,321 \\ 
Adjusted R$^{2}$ & 0.502 & 0.510 & 0.510 & 0.492 \\ 
Residual Std. Error (df = 1315) & 0.706 & 0.700 & 0.700 & 0.713 \\ 
F Statistic (df = 5; 1315) & 267.285$^{***}$ & 275.460$^{***}$ & 275.929$^{***}$ & 256.755$^{***}$ \\ 
\hline 
\hline \\[-1.8ex] 
Standard errors in parentheses. & \multicolumn{4}{l}{$^{*}$p$<$0.1; $^{**}$p$<$0.05; $^{***}$p$<$0.01} \\ 
\end{tabular}
}
\end{table*} 

\begin{table*}[!ht] \centering 
  \caption*{Table S7. Relationship between high output institutions' citation impact and outgoer skill alignment. To allow comparison of the coefficients in the table, all variables were centred. Therefore, the beta coefficients in the table have standard deviations as their units. The results in models (1), (2), (3), and (4) show that the proportion of publications in the top 1\%, 5\%, 10\% have a significant association with the alignment between outgoers and the rest of the workforce. The degree of internal collaboration within institution, PPP(collab$_{\omega})_{i}$, is most strongly associated with alignment within institutions, followed by the total workforce, r$_{i}$, of the institution. Contrary to the regressions for low and medium output institutions, the proportion of outgoers, PP$_{\omega_i}$, in an institution's total workforce is not significantly associated with the outgoers' skill alignment measure.} 
  \label{Table S7} 
{\tsize  
\begin{tabular}{@{\extracolsep{5pt}}lcccc} 
\\[-1.8ex]\hline 
\hline \\[-1.8ex] 
& \multicolumn{4}{c}{\textit{Dependent variable: $\cos{\eta}_{i}$}} \\  
\cline{2-5} 
\\[-1.8ex] & \multicolumn{4}{c}{Institutions with High Publication Output, P$_{i}$ [4864, 121750]} \\ 
\\[-1.8ex] & (1) & (2) & (3) & (4)\\ 
\hline \\[-1.8ex] 
 Constant & 0.584$^{***}$ (0.089) & 0.566$^{***}$ (0.089) & 0.567$^{***}$ (0.088) & 0.840$^{***}$ (0.089) \\ 
  r$_{i}$ & 0.311$^{***}$ (0.023) & 0.297$^{***}$ (0.023) & 0.295$^{***}$ (0.023) & 0.148$^{***}$ (0.052) \\ 
  PP$_{\omega_i}$ & 0.009 (0.024) & 0.009 (0.023) & 0.006 (0.023) & $-$0.018 (0.025) \\ 
  PPP (collab$_\omega)_{i}$ & 0.414$^{***}$ (0.026) & 0.420$^{***}$ (0.025) & 0.420$^{***}$ (0.025) & 0.366$^{***}$ (0.027) \\ 
  PP$_{{top\,1\%}_{i}}$ & 0.276$^{***}$ (0.024) &  &  &  \\ 
  PP$_{{top\,5\%}_{i}}$ &  & 0.297$^{***}$ (0.024) &  &  \\ 
  PP$_{{top\,10\%}_{i}}$ &  &  & 0.299$^{***}$ (0.024) &  \\ 
  TNCS$_{i}$ &  &  &  & 0.270$^{***}$ (0.051) \\ 
 \hline \\[-1.8ex] 
PPS$_{i}$: Diversification/Specialization dummy? & Yes & Yes & Yes & Yes \\ 
\hline \\[-1.8ex] 
Observations & 1,322 & 1,322 & 1,322 & 1,322 \\ 
Adjusted R$^{2}$ & 0.410 & 0.419 & 0.420 & 0.363 \\ 
Residual Std. Error (df = 1316) & 0.768 & 0.762 & 0.761 & 0.798 \\ 
F Statistic (df = 5; 1316) & 184.615$^{***}$ & 191.455$^{***}$ & 192.543$^{***}$ & 151.621$^{***}$ \\ 
\hline 
\hline \\[-1.8ex] 
Standard errors in parentheses. & \multicolumn{4}{l}{$^{*}$p$<$0.1; $^{**}$p$<$0.05; $^{***}$p$<$0.01} \\ 
\end{tabular}
}
\end{table*} 

When analyzing the factors that explain the alignment between outgoers and the rest of the institution, a similar pattern is observed. The results of the analysis are presented in ~\hyperref[Table S5]{Table S5}, ~\hyperref[Table S6]{Table S6}, and ~\hyperref[Table S7]{Table S7}. First, the results in models (1), (2), (3), (4) of the three tables confirm that there is a considerable relationship between an institution's outgoer skill alignment and the proportion of publications in the top 1\%, 5\%, and 10\%, and the total normalized citation score, even after controlling for several other explanatory variables, including the proportion of outgoing researchers, the proportion of internal collaboration between the outgoers and the rest of the institution's workforce, the total workforce, and the degree of diversification of skills. This result is consistent across institutional size subgroups. Interestingly, all tables shows a slightly stronger association between all citation impact measures and `outgoer' skill alignment than for `newcomer' skill alignment for all types of institutions size subgroups. 

Moreover, institutions where researchers collaborate internally are more likely to have converging profiles of outgoing researchers with the rest of the institution. While the proportion of outgoing researchers is related to skill alignment for low and medium output institutions, it is not found to be a significant predictor in high output institutions. In summary, the results suggest that internal collaboration, citation impact, and total workforce capacity are crucial predictors of outgoer skill alignment across different institutional sizes and levels of publication output. These findings suggest that internal collaboration and learning opportunities within institutions may play a major role in aligning the skills of outgoing researchers with the rest of the institution.

\end{document}